\documentclass[aps,prl,reprint,onecolumn,floatfix,
superscriptaddress,nofootinbib,
colorlinks=true,linkcolor=blue,citecolor=blue,urlcolor=blue]{revtex4-2}

\usepackage{graphicx}
\usepackage{amsmath,amssymb}
\usepackage{mathrsfs}  
\usepackage{bm}
\usepackage{mathtools}   
\usepackage{hyperref}
\hypersetup{
	colorlinks=true,
	linkcolor=blue,
	citecolor=blue,
	urlcolor=blue
}

\begin{document}
	
\title{Large-scale array of squeezed light and synchronization using atomic vapor}
\author{Lin Wang}%
\affiliation{State Key Laboratory of Surface Physics, Key Laboratory of Micro- and Nano-Photonic Structures (Ministry of Education) and Department of Physics, Fudan University, Shanghai 200433, China}%
\author{Xichang Zhang}%
\affiliation{National Key Laboratory of Infrared Detection Technologies, Shanghai 200083, China}%
\affiliation{Shanghai Institute of Technical Physics, Chinese Academy of Sciences, Shanghai 200083, China}%
\author{Konstantin Manannikov}%
\affiliation{Department of Physics of Complex Systems, Weizmann Institute of Science, Rehovot 7610001, Israel}%
\author{Nir Davidson}%
\affiliation{Department of Physics of Complex Systems, Weizmann Institute of Science, Rehovot 7610001, Israel}%
\author{Ying Hu}%
\affiliation{State Key Laboratory of Quantum Optics Technologies and Devices, Institute of Laser Spectroscopy, Shanxi University, Taiyuan 030006, China}%
\affiliation{Collaborative Innovation Center of Extreme Optics, Shanxi University, Taiyuan 030006, China}%
\author{Dongdong Hao}%
\email{ddhao19@fudan.edu.cn}
\affiliation{State Key Laboratory of Surface Physics, Key Laboratory of Micro- and Nano-Photonic Structures (Ministry of Education) and Department of Physics, Fudan University, Shanghai 200433, China}%
\author{Yanhong Xiao}%
\email{yxiao@fudan.edu.cn}
\affiliation{State Key Laboratory of Surface Physics, Key Laboratory of Micro- and Nano-Photonic Structures (Ministry of Education) and Department of Physics, Fudan University, Shanghai 200433, China}%
\affiliation{State Key Laboratory of Quantum Optics Technologies and Devices, Institute of Laser Spectroscopy, Shanxi University, Taiyuan 030006, China}%
\affiliation{Collaborative Innovation Center of Extreme Optics, Shanxi University, Taiyuan 030006, China}%

\begin{abstract}
	Quantum light sources such as squeezed light are essential for quantum information science and technologies, but the scalable production of multiple beams of them remains a challenge. Here, we experimentally demonstrate a novel approach to the generation of a large spatial array of polarization-squeezed light beams via atomic-coherence-enhanced nonlinear optical processes using a single atomic vapor cell. Unlike schemes based on independent squeezing generators, the squeezing dynamics of each channel here are governed by a common collective ground-state atomic coherence, produced by all input beams, homogenized by the thermal motion of the atoms, and protected against wall collisions by a paraffin coating. Consequently, the optical states of all channels are coupled and regulated by each other via the moving atoms, leading to synchronization behavior. We realized a 30-beam array of polarization squeezed state with 2.03 $\pm$ 0.02 dB of squeezing, experimentally verified the synchronization, and observed improved purity of the squeezed state as well as the system's response to perturbations when the size of the array increases. This work provides a pathway towards scalable high-performance quantum light sources for applications in precision measurement, quantum imaging and quantum information processing.
\end{abstract}

\maketitle
\section{Introduction}\label{sec1}

Squeezed light is a fundamental resource for quantum networks~\cite{kimble2008quantum, jia2025continuous}, imaging and sensing~\cite{defienne2024advances, zhang2021distributed}, and computing~\cite{deng2023solving, PhysRevLett.97.110501, PhysRevLett.105.053601, 10.1038/nphoton.2015.139, PhysRevA.86.023803, TAYLOR20161, Xia2023}, and there is currently a demand for scalable production of many beams of it. Nowadays, multi-mode squeezed light is mainly generated using optical parametric oscillators (OPOs) with crystals in cavities~\cite{yan2017establishing, zhong2020quantum}, with time- or frequency-multiplexed architectures~\cite{timed, integratedcomb, microcomb}, or multiple setups~\cite{cluster}.  Integrated photonic platforms provide an alternative route toward scalable non-classical light generation~\cite{doi:10.1126/sciadv.adl1814, doi:10.1126/science.abo6213, luo2019nonlinear}, offering compactness and chip-level integration, though predominantly operate near telecommunication wavelengths, and scaling the number of spatial channels remains constrained. Atomic systems constitute a complementary platform for squeezed-light generation~\cite{slusher1985observation}, with the advantage of frequency matching with downstream atomic devices such as quantum memory and quantum sensors~\cite{duan2001long, zhang2011preparation, julsgaard2004experimental}. Multiple quantum optical fields~\cite{PhysRevLett.123.203604, zhang2020reconfigurable, pan2019orbital} from a single atomic vapor cell have been produced using coherence transport~\cite{xiao2006diffusion, PhysRevLett.126.223603, PhysRevLett.123.203604} or multiwave mixing~\cite{PhysRevLett.130.060801}. Recent work has further shown that vacuum vapor cells can yield high squeezing levels in a single beam~\cite{liqing, baerentsen2024squeezed}. Nevertheless, a large-scale spatial array of squeezed light remains unexplored. Beyond the potential for scaling, a shared ground-state atomic coherence across spatial channels introduces an intrinsic coupling among the modes, providing foundation for emergent collective behaviors, such as synchronization.

Synchronization is an intriguing and ubiquitous phenomenon in both natural and artificial systems~\cite{doi:10.1126/science.aav7932}, where coupled oscillators develop aligned dynamical responses through mutual interactions, playing crucial roles in applications such as communication and signal processing. Among the various coupling mechanisms, dissipative coupling, in which oscillators interact dissipatively through a shared environment rather than through direct coherent exchange, provides a particularly robust route to collective behaviour. Owing to its close connection with collective dynamics and correlations, synchronization in quantum systems has recently attracted significant attention and has been explored in a range of platforms, including optomechanical oscillators, trapped ions and quantum oscillators~\cite{PhysRevA.90.033603, PhysRevResearch.5.033209, PhysRevA.109.033718, PhysRevA.99.033818, liu2025observation, li2025experimental}. As a textbook model for oscillators, optical fields such as laser arrays have been found to synchronize in phase through either dissipative or coherent coupling  ~\cite{mahler2020improved, tradonsky2017talbot, doi:10.1126/sciadv.abm7454}. Extending such collective dynamics to quantum optical fields would open a route to robust multi-channel non-classical light generation, with potential for scalable quantum technologies.

In this article, we demonstrate a large-scale spatial array of dissipatively coupled optical channels of squeezed light fields, driven by common collective atomic spin coherence in a vapor cell which can also display synchronization behavior. An array of thirty linearly polarized classical laser beams are all turned into polarization squeezed states with up to $2.03 \pm 0.02$ dB of squeezing, when the orthogonal polarization changes from a coherent vacuum state to a squeezed vacuum upon interacting with the atoms via a coherence-enhanced low-power nonlinear optical process. Due to the thermal atomic motion and the coherence-preserving wall coating of the glass cell, the shared coherence is jointly created by the laser array within the entire cell volume and hence enables the scalability and synchronization of the squeezed light array. We experimentally verified the uniformity in the degree of squeezing, squeezing angle (determined by the phase of the squeezed vacuum) across the array, as well as the synchronization of beams with different laser powers. Furthermore, as the size of the array increases, a higher level of squeezing (before saturation), reduced sensitivity to a defect beam, and enhanced purity of the squeezed state are observed. 

\section{Results}\label{sec2}

\subsection{Theory of squeezed light array}

\begin{figure*}
	\centering
	\includegraphics[width=0.7\textwidth]{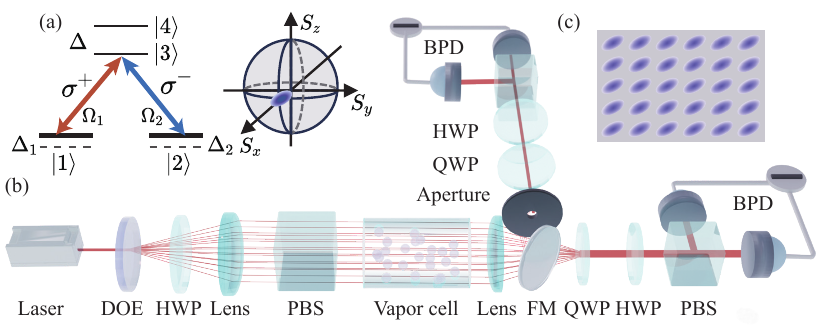}
	\caption{Working principles and experimental schematics of a squeezed light array. 
		(a) Left: Simplified atomic energy levels for squeezing. The linearly polarized light drives the ${}^{87}$Rb $\text{D}_{1}$ transition, resonantly from $|1\rangle$, $|2\rangle$ (belonging to $5S_{1/2}, F=2$ Zeeman manifolds) to $|3\rangle$ ($5P_{1/2}, F'=1$), and off-resonantly to $|4\rangle$ ($5P_{1/2}, F'=2$). Here, $\Omega_{1,2}$ are the Rabi frequencies for the two circularly polarized component respectively, $\Delta=814.5\, \mathrm{MHz}$ is the frequency spacing between the two excited states. $\Delta_{1,2}$ are the AC-stark shifts due to the upper excited state. Right: A Poincar\'{e} sphere illustrating the polarization squeezed state. (b) Experiment setup. A diffractive optical element (DOE) generates 30 spatially separated optical beams propagating in parallel through a coated vapor cell containing ${}^{87}$Rb atoms. The squeezed light array is characterized via two detection schemes: (i) full array quantum fluctuation measurement using balanced homodyne detection, or (ii) single-channel noise power measurement by selecting individual beams with a flip mirror (FM) while blocking others with an aperture. HWP: half-wave plate, PRP: phase-retarding plate, a quarter-wave plate (QWP), PBS: polarization beam splitter, BPD: balanced photo-detectors. (c) An illustration of synchronized squeezed ellipse of the array, where all beams have the same squeezing level and squeezing direction owing to global coupling.} 
	\label{Fig:1}
\end{figure*}

Our light squeezing is based on an established technique called polarization self-rotation (PSR)~\cite{PhysRevA.66.043815, mikhailov2008low, ries2003experimental, liqing}, which can be described by a simplified four-level atom interacting with a linearly polarized (along $x$) laser beam. As shown in Fig.~\ref{Fig:1}(a), a $\Lambda$-type electromagnetically induced transparency (EIT)~\cite{novikova2012electromagnetically} is formed between an excited state $|3\rangle$ and two ground state degenerate Zeeman levels $|1\rangle$ and $|2\rangle$. With a frequency difference $\Delta$ to $|3\rangle$, the upper excited state $|4\rangle$ induces AC-stark shift $\Delta_{1,2}=\Omega_{1,2}^2/\Delta$ to $|1\rangle$ and $|2\rangle$ respectively, where $\Omega_{1,2}$ is the Rabi frequency of the $\sigma^{+,-}$ components of the input laser. This causes a polarization rotation denoted by the light's Stokes vector component $S_y$~\cite{supplementary2025} proportional to $S_{z}=(\Omega_{1}^{2}-\Omega_{2}^{2})/(\Omega_{1}^{2}+\Omega_{2}^{2})$, which describes a one-axis-twisting (OAT) dynamic $S_{z}^{2}$ for the light spin $S$ on the Poincar\'{e} sphere~\cite{ries2003experimental,PhysRevA.91.051804}, and leads to shearing and squeezing of the uncertainty ellipse that determines both the squeezing level and the squeezing angle, or vacuum squeezing in the orthogonal $y$-polarization. We emphasize that the polarization rotation is enhanced by the ground state atomic coherence (due to EIT) which depends on the laser power. This process can be also understood as a degenerate four-wave-mixing (FWM) process, where the $x$-polarized input (pump) field couples the two orthogonal superposition ground states $|1\rangle+|2\rangle$, $|1\rangle-|2\rangle$ to $|3\rangle$ and $|4\rangle$ respectively, and produce a pair of quantum-correlated $y$-polarized photons. This type of squeezed light has been studied extensively in vapor cell systems, mostly with a single squeezed light beam~\cite{PhysRevA.93.013853, PhysRevA.96.013835} and also in the context of quantum-enhanced atomic magnetometers~\cite{PhysRevA.91.051804, zhang2021dichroism, PhysRevA.86.023803}. 

In the current system of a large laser array interacting with warm atoms in a coated vapor cell, squeezing of each optical channel is enhanced or even dominated by the common ground-state atomic coherence. This coherence is produced collectively by all laser beams and homogenized by the fast motion of the atoms, where the key is that the paraffin coating on the inner wall of the vapor cell~\cite{PhysRevA.66.042903} protects the coherence against thousands of atom-wall collisions. The collective spin state lowers the laser power threshold per optical channel for squeezing and synchronizes the phases and the squeezing levels of all the squeezed vacuum.  To describe the array dynamics, we treat each illuminated channel as a local double-$\Lambda$ atom-light interaction region, while the dark region un-illuminated by light acts as a shared and dynamically-evolving reservoir of long-lived atomic coherence. Atoms acquire ground-state coherence in one optical channel, then leave the beam, travel and evolve in the dark region, and later randomly re-enter the array. The inter-channel coupling is therefore global and dissipative. In the linear polarization basis of the laser propogating along $z$-axis, the atom-light interaction in the $i$-th channel is
\begin{equation}
	\begin{aligned}
		\hat{H}_{int} &= \frac{\hbar N_{i}}{L} \int_{0}^{L} dz \big[ \delta' \hat{\rho}_{33} + (\delta' + \Delta) \hat{\rho}_{44} - g_2 \alpha \hat{\rho}_{3y} - g_1 \alpha \hat{\rho}_{4x} 
		- g_2 \hat{a}_y \hat{\rho}_{3x} - g_1 \hat{a}_y \hat{\rho}_{4y} + \mathrm{H.c.} \big],
	\end{aligned}
\end{equation} 
where $\hbar$ is the reduced Planck constant, $N_i$ is the atom number of $i$-th channel, $L$ is the length of the atomic medium, $g_{1,2}$ are the single-photon coupling strengths, $\alpha$ is the amplitude of the $x$-polarized drive, $\delta'$ is the one-photon detuning, and $\Delta$ is the excited-state hyperfine splitting. The coupled atomic dynamics are described by
\begin{equation}
	\begin{aligned}
		\dot{\hat{\rho}}^{(i)} &= -\frac{i}{\hbar}[\hat{H}_{int}^{(i)}, \hat{\rho}^{(i)}] + \hat{\mathscr{L}}^{(i)}[\hat{\rho}^{(i)}]- k_{i0}\hat{\rho}^{(i)} + k_{i0}\hat{\rho}^{(0)} + \hat{\mathfrak{F}}_i, 
	\end{aligned}
\end{equation}
\begin{equation}
	\begin{aligned}
		\dot{\hat{\rho}}^{(0)} &= -\gamma_{0}\hat{\rho}^{(0)} + \hat{\mathscr{L}}^{(0)}[\hat{\rho}^{(0)}]+ \sum_{i=1}^{N}k_{0i}\hat{\rho}^{(0)} - \sum_{i=1}^{N}k_{0i}\hat{\rho}^{(i)} + \hat{\mathfrak{F}}_0 ,
	\end{aligned}
\end{equation} 
where $\hat{\rho}^{(i)}$ and $\hat{\rho}^{(0)}$ denote the atomic density matrices in the optical channels and in the dark region, respectively, $\hat{\mathscr{L}}^{(i)}$ includes relaxation and repopulation processes, $k_{i0}= \frac{dN_{i}}{N_idt}$ and $k_{0i}= \frac{dN_i}{N_ddt}$ are the exchange rates between channel $i$ and the reservoir, where $N_d$ is the atom number in the dark region. $\gamma_0$ is the decay rate of the ground state population, and $\hat{\mathfrak{F}}_{i,0}$ are Langevin noise operators. The generated $y$-polarized quantum field obeys
\begin{equation}
	\begin{aligned}
		\left( \frac{\partial}{\partial t} + c\,\frac{\partial}{\partial z} \right) 
		\hat{a}_{y}^{(i)}(z, t) &= i g_{1} N_{i}\, \hat{\rho}_{y4}^{(i)} + i g_{2} N_{i}\, \hat{\rho}_{x3}^{(i)}.
	\end{aligned}
\end{equation}
This framework shows that the effective inter-channel coupling is set by the channel-reservoir exchange rates and the atomic coherence lifetime in the dark region, which depends on the atomic density and the geometry of the laser beam and vapor cell. The channel spacing plays negligible roles in this global coupling. Our numerical model \cite{supplementary2025} incorporates spontaneous emission, Doppler broadening, coherence decay and population exchange with the lower hyperfine energy level, and reproduces the experimental observations qualitatively. We next present the main experimental results obtained from the setup shown schematically in Fig.~\ref{Fig:1}(b), with more experimental details provided in the Materials and methods.

\subsection{Performance of a multiplexed squeezed light array}

\begin{figure*}
	\centering
	\includegraphics[width=0.7\textwidth]{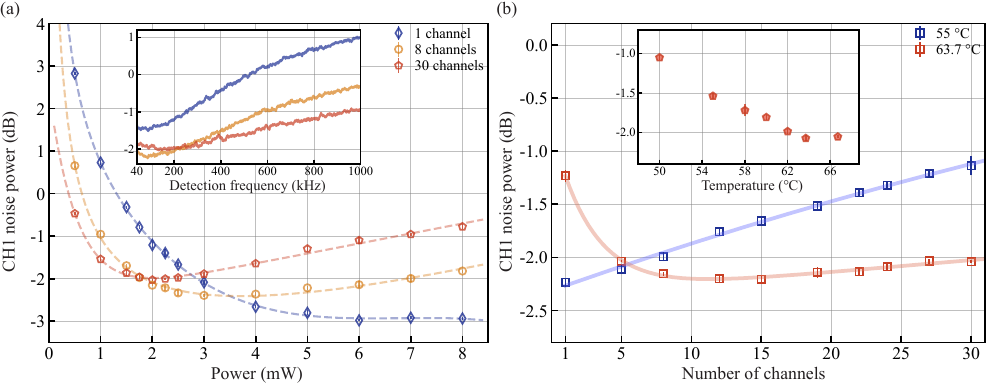}
	\caption{Experimental results of squeezed array.
		(a) Noise power of the minimal-noise-quadrature (denoted as squeezed quadrature) of a selected channel (CH1) relative to the photon shot noise level (set as $0$ dB) versus the laser power per channel, for a single channel case, a 8-channel and 30-channel array respectively, while the beam size of all channel is kept at $0.505$ mm in all cases. The cell temperature is $63.7^{\circ}$C. The dashed curve is a fit to guide the eye. (b) Squeezed quadrature noise of CH1 versus the array size at two temperatures ($55^{\circ}$C and $63.7^{\circ}$C), with a fixed input laser power of $2$ mW and a corresponding beam size of $0.505$ mm for each channel. The inset shows the best squeezing achievable at different cell temperatures for a 30-channel array. The corresponding optimal power per channel for maximal squeezing is $[0.75\,\mathrm{mW}, 1\,\mathrm{mW}, 1.25\,\mathrm{mW}, 1.5\,\mathrm{mW}, 1.75\,\mathrm{mW}, 2\,\mathrm{mW}, 2\,\mathrm{mW}]$ for the temperatures of $[50^{\circ}\text{C}, 55^{\circ}\text{C}, 58^{\circ}\text{C}, 61^{\circ}\text{C}, 62^{\circ}\text{C}, 63.7^{\circ}\text{C}, 66.7^{\circ}\text{C}]$ respectively. The solid curve is a fit to guide the eye. The shot-noise-limited performance of our detection system and laser source was verified by measuring the linear dependence of the noise power on the optical power. The shot noise level exceeded the electronic noise by approximately $20$ dB within our measurement bandwidth. All noise powers are obtained at a detection frequency of $180$ kHz. Spectrum analyzer settings are: resolution bandwidth (RBW) = $10$ kHz, with each noise spectrum averaged for at least 40 times. Each data point represents the mean of six or more repeated measurements, and the error bars denote the standard deviation of these values.
	}
	\label{Fig:2}
\end{figure*}

The simultaneous generation of squeezed light in a multiplexed array of $30$ beams constitutes a key result of our work. We proceed to its detailed characterization by presenting the data from an arbitrarily selected channel (named CH1) as representative, with consistent performance observed across all channels. As depicted in Fig.~\ref{Fig:2}(a), the measured noise power for the squeezed quadrature in the 30-channel array is $2.03$ dB below the photon shot noise level (set as the $0$ dB reference) for a laser power of $2$ mW per channel. We emphasize that, for laser power (per channel) of about $1$ mW, there is no squeezing for the single-channel case, but about $1.5$ dB for the 30-beam array. Even at a power as low as $0.5$ mW per channel, the array generates squeezing of about $0.47$ dB, whereas the single-channel case shows excessive noise of about $3$ dB above the shot noise even for the minimal-noise quadrature. These results show that one can achieve squeezing in the entire array with a relatively low laser power per channel which for one channel by itself is insufficient to generate squeezing. In this regime, the system's performance is dominated by the collective dynamics of the entire array, rather than by independent processes in individual channels. Therefore, to produce multiple squeezed beams, using a spatial array in one coated cell is superior to employing multiple vapor cells each with a single-beam, because of the better scalability and lower overall laser power. 

For equal power per channel, larger arrays sustain squeezing to higher frequencies. As shown in the inset of Fig.~\ref{Fig:2}(a), for $2$ mW incident power per channel, a representative channel (named CH1) in the 30-channel array still exhibits about $1$ dB of squeezing at $\sim$900 kHz and an effective bandwidth of $\sim$890 kHz, compared with $\sim$283 kHz for the single-channel configuration. This behavior arises from the larger total optical power in the cell, which enhances the collectively mediated atomic coherence that drives four-wave mixing. This should, however, be distinguished from the fixed-total-power comparison between arrays with different channel numbers, where distributing the same total power over more channels reduces the local pump intensity and hence both the squeezing level and bandwidth become smaller in an individual channel.

The observed achievable squeezing, however, saturates and eventually degrades for higher laser power, revealing the performance of such quantum arrays is ultimately constrained by the finite optical depth. As in any nonlinear optical process, higher power is often needed for larger nonlinearity. Here, with the assistance of atomic coherence, relatively lower laser power is required but saturation is unavoidable. As shown in Fig.~\ref{Fig:2}(a) the squeezing level initially increases with laser power, and then plateaus and worsens. Moreover, compared to smaller arrays, the 30-channel larger array saturates at a lower power (per channel) and the achievable maximal squeezing is less. For instance, the 1-channel configuration can attain about $3$ dB of squeezing (at $6$ mW), while the 8-channel and the 30-channel arrays only reach $2.39$ dB (at $3$ mW) and $2.03$ dB (at $2$ mW) respectively.  While increasing the total laser power in the array enhances the atomic coherence within the cell, it also introduces competing effects that limit the achievable squeezing. In the near-resonant regime, strong optical pumping redistributes population among the hyperfine ground states, transferring atoms into the $F=1$ that do not participate in the four-wave mixing process. This population leakage lowers the effective optical depth and the four-wave mixing gain available for squeezing generation as the array size increases, leading to the observed saturation in larger arrays. At the same time, stronger resonant interaction increases spontaneous-emission-induced noise, imposing an additional limit on the achievable squeezing. 

Consistent with this picture, increasing the atomic density partially compensates the reduced effective optical depth. As shown in Fig.~\ref{Fig:2}(b), for lower atomic density, for example at $T = 55^{\circ}$C, at a fixed $2$ mW laser power per channel, squeezing is monotonically worse with increased channel number, while for $T=63.7^{\circ}$C, squeezing only becomes slightly worse for more than 15 channels. Furthermore, the inset shows the overall maximal squeezing level for a 30-channel array increases with vapor cell temperatures. However, further heating in our system is prohibited due to degradation of the paraffin coating, and the expected improvement at higher vapor temperatures under thermally more stable wall-coatings are theoretically analyzed in \cite{supplementary2025}. These results highlight a trade-off between mode number and per-mode squeezing in the present architecture, which prioritizes hardware-efficient generation of many spatially distinct squeezed channels from a single atomic cell, complementing cavity-based OPO and integrated photonic approaches.

\subsection{Synchronization of the squeezed light array}

\begin{figure*}
	\centering
	\includegraphics[width=0.7\textwidth]{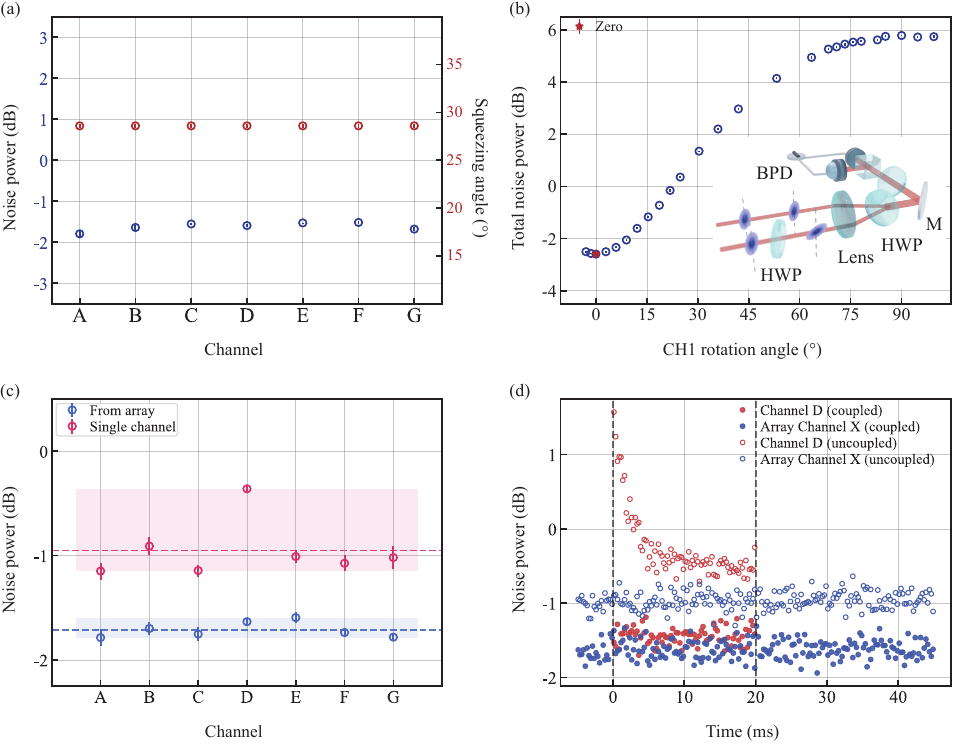}
	\caption{Synchronization in the squeezed light array. (a) The squeezed quadrature noise (blue) and squeezing angle (red) of seven randomly selected channels in a 30-channel array. Laser power $2$ mW and beam size $0.505$ mm for each channel. (b) The inset in (b) illustrates the experimental schematic at the output of the cell for measuring the total noise power using a two-channel array. One of the two squeezed beams passes through an additional quarter-wave plate (QWP), which rotates its polarization ellipse before combining with the other beam. The main panel shows the total noise power, measured via balanced homodyne detection, as a function of the rotation angle applied by the QWP. Both input squeezed beams have an optical power of $5$ mW. The red pentagram denotes the noise level without rotation, while the blue points show the increase in total noise power with varying rotation angle. (c) An extra channel (Channel D) is added to the 30-channel array, where each channel of the array has a laser power of $2$ mW. The laser power of Channel D is set to $1.5$ mW, with a beam size identical to that of the other array channels. We measured the degree of squeezing of six channels from the array (A, B, C, E, F, G) and Channel D, under two conditions: single-channel operation case (uncoupled regime), and full 31-beam array operation case (coupled regime). The detection frequency is $186$ kHz. Error bars (vertical lines in the symbol) represent standard deviations from six repetitive measurements. (d) Synchronization dynamics. Channel D, with a laser power of 1.5 mW, is operated as a pulsed beam, with the pulse starting at 0 ms and ending at 20 ms, as indicated by the black vertical dashed lines. One channel (named channel X) selected from the 30-channel continuous-wave light array ( $2$ mW per channel) is monitored for comparison. We measured the time evolution of the degree of squeezing for both Channel D and Channel X under two conditions: single-channel operation (uncoupled regime) and full 31-beam array operation (coupled regime). The homodyne output is fed into a lock-in amplifier and demodulated at 230 kHz. The variance of the demodulated amplitude is evaluated as a function of time and used to extract the time-dependent squeezing. Channel D is pulsed at 20 Hz, and the acquisition is synchronized with the pulse trigger.}
	\label{Fig:3}
\end{figure*}

Having established the squeezed array, we now explore the uniformity and synchronization in the squeezing properties, enabled by flying atoms acting as a global coupling medium inherently linking all channels within the array. We point out that, the phase of the $y$-polarized squeezed vacuum relative to that of the $x$-polarized input laser determines the orientation of the squeezed ellipse on the Poincar\'{e} sphere, or the squeezing direction. Since the phase of laser array from the DOE is the same, as experimentally proved by a far field image~\cite{supplementary2025}, we can conclude that the squeezed vacuum array are phase synchronized if the squeezing angle is the same across the array. We provide experiment results to verify the uniform squeezing angle of the generated squeezed array. The squeezing angle, denoted by the direction $\theta$ of the long axis of the ellipse on the Poincar\'{e} sphere relative to the equator plane, is derived and calibrated from the phase retarder’s rotation angle~\cite{supplementary2025}. For seven randomly selected channels, nearly identical squeezing angles and squeezing level are observed, as shown in Fig.~\ref{Fig:3}(a). This measurement of selecting an individual element from the array is however affected by loss due to the clipping by the aperture when selecting a certain beam (Fig.~\ref{Fig:1}(b)).  

To further prove the uniform squeezing direction, we perform additional experiments where the squeezing level of the array as a whole is detected and compared with a single channel. As shown in Fig.~\ref{Fig:3}(b) inset, we first perform a two-channel array experiment where for one channel the phase of the $y$-polarized squeezed vacuum field is shifted via a phase retarder (QWP) before the lens, thus rotating its squeezing ellipse. The detected squeezing of the total array shown in Fig.~\ref{Fig:3}(b), changes with this rotation in an expected trend consistent with our theoretical simulations~\cite{supplementary2025}: the overall squeezing is maximal when the two ellipses are aligned, with a monotonic degradation as their angular difference grows.

Then, we perform such experiments as in Fig.~\ref{Fig:3}(b) for an array of 10 channels (laser power $1$ mW per channel), but respectively let 10, 5, and 1 channels into the detector system (without mis-aligning them as above), and observe nearly identical squeezing of $1.18$ dB, $1.19$ dB and $1.19$ dB respectively. Because of laser power limit of the photo-detectors we could not examine a larger array. These experiments prove that all the squeezed quadratures are oriented along the same direction. 

Synchronization often manifests itself as unifying the initially different phases or amplitudes of many oscillators by introducing coupling between them. In our system, when channels with different laser power give different squeezing levels for the independent single-channel operation, their squeezing can be synchronized in the array operation. The key signature of the regime for synchronization here is when the squeezing level versus the laser power exhibits a plateau as shown in Fig.~\ref{Fig:2}(a). This plateau indicates saturation, similar to the saturation regime in classical mode-locked lasers where the system settles into a globally stable synchronized state. To verify this, we introduce an additional channel (named ``D") with $1.5$ mW laser power in the 30-channel array uniform array with $2$ mW per channel, and we measure squeezing of six channels (A,B,C,E,F,G) from the array and the channel D. As seen in Fig.~\ref{Fig:3}(c), when only one channel is present in the vapor cell, channel D shows less squeezing than others, which is consistent with the single-channel case in Fig.~\ref{Fig:2}(a). When all the 31 beams are in the cell, channel D has similar squeezing with the rest while the squeezing in all channels are enhanced. As indicated by the colored bar Fig.~\ref{Fig:3}(c), the inter-channel difference in the squeezing level is reduced from 0.79 dB to 0.19 dB when the system changes from a uncoupled to a couple regime.

We emphasize that this synchronization behavior in the saturation regime is analogous to that in both mode-locked lasers~\cite{Yoshita_2009,Chen_2024} and phase-locked laser arrays~\cite{mahler2020improved} where the synchronized mode survives because it suffers minimal loss by the saturable absorber. The synchronized state is dynamically selected out among all possible modes. In passively mode-locked lasers, the saturable absorber provides the nonlinear loss discrimination, where the high-peak-power pulse mode experiences minimal loss, thereby strongly suppressing the continuous wave mode. In coupled laser arrays~\cite{mahler2020improved}, the phase-locked mode corresponds to the highest total laser power focused on the saturable absorber in a cavity, also corresponding to minimal loss. In both cases, the saturation mechanism provides the critical nonlinear process required for the emergence of the synchronized mode. Here the entire atomic medium plays the role of the saturable absorber because the atoms are flying and the entire array see the same atomic ensemble, which is equivalent to focusing the whole laser array into a solid-state absorber as in ~\cite{saturable}. In essence, synchronization occurs when all the 31 squeezed vacuum fields are in phase and the forward scattered $y$-polarized quantum light is maximal in total (hence squeezing is maximal), which is resulted from the long-lived (minimal loss) Doppler-free collective spin state that survives. 

We note that, although in principle, synchronization could also manifest through the squeezing angle, in practice due to spontaneous emission noises the squeezed state combines modest squeezing with large excessive noises especially in the anti-squeezed quadrature. This effectively smears out the uncertainty ellipse and renders the orientation of the noise ellipse insensitive to small variations induced by laser power changes~\cite{supplementary2025}. As a result, when experimentally validating synchronization, we can only measure the squeezing levels, which are inherently connected to the squeezing angle in the OAT light squeezing scheme. Direct experimental proof of dynamic phase synchronization or locking in the squeezed vacuum array is not feasible in our experiment. 

To further substantiate synchronization, we probe the temporal dynamics through a pulsed experiment as shown in Fig.~\ref{Fig:3}(d). We applied a localized pulsed pump of 20 ms in duration with laser power of 1.5 mW to an optical channel (named Channel D) outside of the 30-beam array where each channel has input laser power of 2 mW. We then monitored the time evolution of the squeezing level of Ch.D, and a channel (named Channel X) from the array for comparison. For the uncoupled case, i.e., when Ch.D (or Ch.X) is present alone in the Rb cell, one can observe the establishment of the steady state near the turn-on time $t=0$ for Ch.D, which takes about 2.7 ms, and that the squeezing level of $0.5$ dB is different than the $1$ dB of Ch.X because their laser power is different. In contrast, for the coupled case, i.e., when Ch.D and Ch.X coexist with the rest of the array and they are all coupled, Ch.D’s squeezing rapidly reaches about the same level ($1.5$ dB) of the array with a nearly invisible transit near $t=0$, which is a quick synchronization. The difference in the transit time for Ch.D in the coupled and uncoupled cases is due to the following. The dissipation rate of the collective coherence determines the time for the system to reach steady state upon perturbation, and this is determined by both the physical movement of the atoms containing many wall bounces (i.e., coherence transport) and the collective atom-light interaction, i.e., the same set of atoms interact with all the 30 or 31 optical channels which bring nonlinearity and saturation. The total laser power in the vapor cell for the 31-beam array (coupled regime) is much higher than for the 1-beam case (uncoupled regime), and hence larger array causes faster optical pumping process and therefore quicker dynamics. 

\subsection{Study of the array's response to imperfection}

\begin{figure*}
	\centering
	\includegraphics[width=0.7\textwidth]{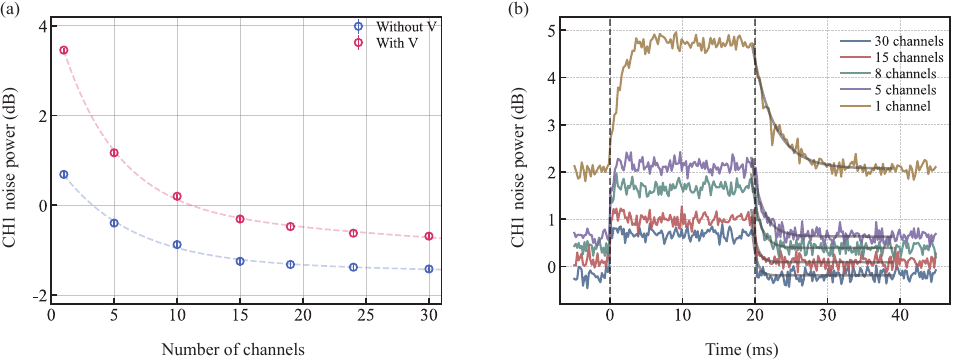}
	\caption{Response of the array to a defective channel. (a) Degradation of squeezing due to a ``defect" channel with input light polarization (V) orthogonal to that of the array, for arrays containing different numbers of channels. The solid curve is a fit to guide the eye. The defect beam: $2$ mW, waist $0.525$ mm. The H-polarized beams in the array: $1$ mW per channel, waist $0.505$ mm. The detection frequency is $186$ kHz. Error bars (vertical lines in the symbol) represent standard deviations from six repetitive measurements. (b) Time-resolved response of the array to a pulsed defect channel. The defect beam has a power of 1.0 mW and a waist of 0.468 mm, whereas the H-polarized array beams have a power of 0.5 mW per channel and a waist of 0.505 mm. The defect beam is switched on at 0 ms and switched off at 20 ms, while the H-polarized array beams remains in continuous wave operation. The squeezing level of one selected H-polarized channel in the array is recorded as a function of time using the same acquisition method as in Fig.~\ref{Fig:3}(d). Arrays containing 1, 5, 8, 15 and 30 channels are measured, yielding recovery times of $2.99\pm 0.09, \, 1.54\pm 0.08,\, 1.11\pm 0.05,\, 0.67\pm 0.06$ and $0.59\pm 0.07$ ms respectively (with errors from fitting uncertainties), after removal of the defect channel. }
	\label{Fig:4}
\end{figure*}

Furthermore, we investigate the response of the squeezed array to imperfections, i.e., the array's ability to mitigate the detrimental effect of a ``defect" optical channel, and found that a larger array is more tolerant than a smaller one. For an array containing just two optical channels with orthogonal polarizations (otherwise with identical parameters), the coherence generated by the two channels interferes destructively~\cite{peng2016anti}, thus suppressing light squeezing in both channels. Crucially, when this orthogonally polarized ``defect" beam is introduced into an array, the resulting modification of the atomic spin state and destructive interference propagate through the coupled medium. This leads to a system-wide breakdown of the required ground-state atomic coherence across all channels. This global suppression of coherence is the underlying mechanism for the observed rapid degradation of squeezing throughout the entire array, strongly suggesting that the quantum noise reduction relies on a collective, highly coherent state established by all coupled channels. However, as the channel number in the array (all with horizontal (``H") polarizations) increases, the detrimental effect of the ``defect" vertically (``V") polarized channel is reduced through motional averaging of the atoms, causing the net coherence to asymptotically approach that of a pure H-polarized array and some recovery of the squeezing. Nevertheless, the additional quantum noise introduced by the V-channel cannot be averaged out. As shown in Fig.~\ref{Fig:4}(a), while the degradation of H-channel squeezing due to the defect V-channel is alleviated with increasing channel number (all ``H"), a residual damage remains, preventing a full recovery of squeezing. 

The recovery dynamics after removal of the defect are characterized in an independent time-resolved measurement in Fig.~\ref{Fig:4}(b). This measurement is carried out at lower optical powers, because the dynamics become too fast to visualize at higher powers as explained above.  As can be seen, the squeezing of a monitored channel from the array (named CH1) gradually recovers once the defect pulse is switched off. The recovery time decreases from $2.99\pm 0.09$ ms in the single-channel array case to $0.59\pm 0.07$ ms for a 30-channel array, confirming the previous analysis that larger arrays rebuild the collective coherent state faster than samller array after a perturbation.

\subsection{The purity of squeezed light state}

\begin{figure*}
	\centering
	\includegraphics[width=0.7\textwidth]{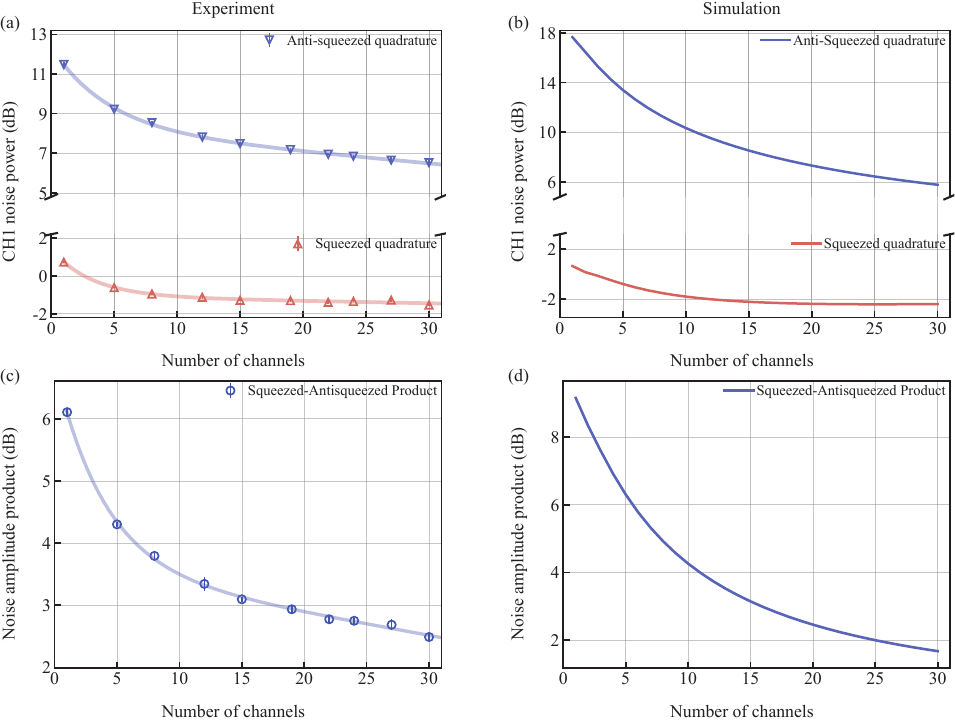}
	\caption{Improved purity of squeezed states for a larger array. Experimental and numerical results of the squeezed and anti-squeezed quadrature noises of a selected channel (CH1) in the array, for arrays containing different number of channels. The solid curve in (a) is a fit to guide the eye. Laser power of $1$ mW and beam size of $0.505$ mm for each channel. Detection frequency $180$ kHz. Spectrum analyzer settings are the same as that in Fig.2. Error bars (small vertical lines inside the symbols) represent standard deviations from six repetitive measurements. Panels (c) and (d) show, in dB, the products of the squeezed and anti-squeezed noise amplitudes, equal to the mean of the two dB numbers in the upper panel (a) and (b) respectively. Zero dB corresponds to a pure unitary squeezed state. Despite the qualitative agreement between the theory and experiment, some quantitative deviation remains mainly due to neglecting experimental imperfections in the simplified model, such as inhomogeneities in the laser powers and polarizations in the array beams, optical losses, finite detection efficiency, and residual magnetic field inhomogeneity and noises etc.}
	\label{Fig:5}
\end{figure*}

Finally, an interesting feature of our quantum light array is that the purity or unitarity of the squeezed light can be manipulated by leveraging the balance between squeezing and saturation. The Heisenberg uncertainty principle dictates that an ideal unitary squeezing process produces the same amount of squeezing and anti-squeezing, a property critical in harnessing quantum resources ~\cite{PhysRevA.87.043826, braverman2018impact}. However, excessive anti-squeezing often appears due to dissipation processes which degrades the entanglement. Here, light scattering during squeezing~\cite{PhysRevA.73.023806} induces excessive noise above the photon shot noise level. As shown in Fig.~\ref{Fig:5} this is most severe when the total laser power is low, rendering a relatively weak ground-state coherence and a negligible EIT effect. Remarkably, when the channel number increases, the initially high noise level in the anti-squeezed quadrature drastically decreases, indicating improved squeezed-state purity. This is mainly attributed to population loss to $F=1$ states at higher total laser power in the array which reduces the absorption and associated noises. These experimental results are in good qualitative agreement with theoretical simulations (detailed in~\cite{supplementary2025}) in Fig.~\ref{Fig:5}(b). To quantify the state purity, we also plot in Fig.~\ref{Fig:5}(c) and (d) the product of the squeezed and anti-squeezed noise amplitudes, derived from the square root of the corresponding noise variances. The product decreases with increasing channel number, indicating reduced excess noise and improved purity of the squeezed states.

\section{Discussions}\label{sec3}

In conclusion, we have demonstrated a large-scale laser array of polarization squeezed states using a rubidium vapor cell. The squeezing level and angle of all the beams in the array are aligned due to the collectively built and shared atomic coherence by the laser array that drive the nonlinear optical process for squeezing. Synchronization behavior can be observed for moderately mismatched channels. The path to higher squeezing in our system includes using a longer cell, and a coating more resistant to higher temperatures~\cite{zhang2015effects, liao2015time} with proper optimization on laser power and laser detunings, to obtain larger optical depth while maintaining the atomic coherence. Using a wider cell also helps alleviate saturation~\cite{supplementary2025} hence improves squeezing (see \cite{supplementary2025} for quantitative analysis). Another route is to incorporating a different underlying squeezing mechanism based on off-resonant atom-light interaction in presence of a magnetic field that gives light squeezing as high as about 10 dB~\cite{baerentsen2024squeezed} and is compatible with our diffusively-coupled array protocol. Our approach to large-scale squeezed light array has the advantage of scalability, synchronization, tunability, tolerance to system imperfections and compatibility with downstream atomic devices. 

When further improvements are implemented, this architecture could be used for spatially resolved quantum-enhanced sensing, quantum imaging, optical computing and quantum networks. In particular, although the beams in the array in our experiment are not quantum-correlated to each other~\cite{PhysRevLett.126.223603}, through followed linear optical processing~\cite{Taylor2013, PhysRevLett.98.083602}, multipartite entanglement states can be created, and be subsequently used for quantum-enhanced distributed sensing networks~\cite{2hsx-5qfr, guo2020distributed} or multi-parameter sensing~\cite{li2026quantum, ma2026high}.  

\section{Materials and methods}\label{sec4}

\subsection{Experimental setup}

The experimental setup is designed to generate and characterize a multiplexed array of squeezed light beams as illustrated in Fig.~\ref{Fig:1}(b). A laser, tuned to the ${}^{87}$Rb $\text{D}_{1}$ $F=2 \rightarrow F'=1$ transition (795 nm), is split into a two-dimensional 7×7 array by a custom diffractive optical element (DOE) with a measured intensity uniformity of $\sim 91\%$. Each beam has a waist ($1/e^{2}$) of $0.505$ mm at the cell center, with single-beam powers ranging from $0.5$ to $8$ mW ($2$ mW yields optimal squeezing for 30-channel array). The $3$ mm inter-beam spacing is preserved over $0.75$ m with negligible distortion. The atomic ensemble is housed in a cylindrical glass cell ($2.5$ cm diameter, $7.5$ cm length) filled with enriched ${}^{87}$Rb. To preserve ground-state coherence, the cell inner wall is coated by paraffin that shows no degradation over years of operation below its $67^{\circ}$C melting point. Experiments are conducted at an optimal temperature of $63.7^{\circ}$C. The cell is placed inside a four-layer magnetic shield. Because the vapor cell is not positioned exactly at the geometric center of the four-layer magnetic shield, and the $2.5$ cm-wide apertures of the shield layers are not perfectly concentric, the outer ring of the array is partially clipped, leaving only $30$ fully transmitted optical channels. The noise power of the output array, with each beam consisting of the $x$-polarized pump and the generated squeezed vacuum, is measured via balanced homodyne detection (BHD) system, where the pump field itself acts as the local oscillator\cite{PhysRevA.91.053848}. Two BHD systems are employed and switched via a flip mirror. To quantify the light squeezing of a beam, one BHD was used to directly measure individual channels. To probe synchronization across the array, a second BHD was configured with a lens to focus and measure a group of beams simultaneously. Details of the BHD configuration are provided in Supplementary Note.

\section*{Data Availability}

All the data supporting the findings of this study are available within this article and its Supplementary Information. Any additional information can be obtained from corresponding authors on reasonable request.

\section*{Acknowledgements}

We thank Irina Novikova and Eugeniy E. Mikhailov for helpful discussions. This work is supported by the Innovation Program for Quantum Science and Technology under Grant No. 2023ZD0300900, STCSM24LZ1400400, the Natural Science Foundation of China (NSFC) under Grants No. 12161141018, and No. 12027806, and the Key Project of the National Natural Science Foundation of China Joint Funds (Grants No. U25A20197).

\section*{Competing Interests}

The authors declare no competing interests.

\section*{Supplementary Information}

\section*{Supplementary Note 1: One-axis-twisting light squeezing by polarization self-rotation}

The polarization self-rotation (PSR) effect can be described using a double-$\Lambda$ four-level system, where the linearly polarized light couples the ground states ($|1\rangle$, $|2\rangle$) and the excited states ($|3\rangle$, $|4\rangle$). To simplify the model, we assume that the energy splitting $\Delta$ between $|3\rangle$ and $|4\rangle$ is large, so that when the laser is near resonant with $|3\rangle$, the influence of $|4\rangle$ is solely AC-Stark shifts of the ground states. Consequently, in the circular polarization basis, the Hamiltonian of this system takes the simplified form:

\begin{equation}
	H_{int} = \frac{\hbar N_{0}}{L} \int_{0}^{L} dz \left[ \Delta_{1} \hat{\rho}_{11} + \Delta_{2} \hat{\rho}_{22} - \Omega_{1}\hat{\rho}_{13} - \Omega_{2} \hat{\rho}_{23} + \text{H.c.} \right],
\end{equation} where $N_{0}$ is the number of atoms and $L$ is the length of the atomic medium. The Rabi frequencies for the right- and left-circularly polarized pump light are $\Omega_{1}=\frac{d_{R}E_{R}}{\hbar}$ and $\Omega_{2}=\frac{d_{L}E_{L}}{\hbar}$ respectively, with $d_{R,L}$ being the dipole moments. The light-induced AC Stark shifts are given by $\Delta_{1}=\Omega_{1}^{2}/\Delta$ and $\Delta_{2}=\Omega_{2}^{2}/\Delta$. For $x$-polarized light, $\Omega_{1}=\Omega_{2}=\Omega$, and the relative phase between the right and left circular polarization components is zero,  $\Omega_{1}^*=\Omega_{1}, \Omega_{2}^*=\Omega_{2}$. The system dynamics is governed by the master equation:

\begin{equation}
	\begin{cases}
		\dot{\rho}_{11}=-\gamma_0 \rho_{11} + i \Omega_1 \rho_{13} + \gamma_0 \rho_{22} - i \Omega_1 \rho_{31} + \gamma \rho_{33} \\
		\dot{\rho}_{12}=-\gamma_{12} \rho_{12} + i \Delta_1 \rho_{12} - i \Delta_2 \rho_{12} + i \Omega_2 \rho_{13} - i \Omega_1 \rho_{32} \\
		\dot{\rho}_{13}=i \Omega_1 \rho_{11} + i \Omega_2 \rho_{12} - \gamma \rho_{13} + i \Delta_1 \rho_{13} - i \Omega_1 \rho_{33} \\
		\dot{\rho}_{22}=\gamma_0 \rho_{11} - \gamma_0 \rho_{22} + i \Omega_2 \rho_{23} - i \Omega_2 \rho_{32} + \gamma \rho_{33} \\
		\dot{\rho}_{23}=i \Omega_1 \rho_{21} + i \Omega_2 \rho_{22} - \gamma \rho_{23} + i \Delta_2 \rho_{23} - i \Omega_2 \rho_{33} \\
		\dot{\rho}_{33}=-i \Omega_1 \rho_{13} - i \Omega_2 \rho_{23} + i \Omega_1 \rho_{31} + i \Omega_2 \rho_{32} - 2\gamma \rho_{33}
	\end{cases}
\end{equation} 
where $\gamma_{0}$ is the decay rate of the ground state population, $\gamma$ is the decay rate of optical coherence, and $\gamma_{12}$ is the decay rate of ground-state coherence. The steady-state solution of the optical coherence can be derived:

\begin{equation}
	\rho_{13} = \frac{\frac{i}{2}\Omega_{1}+i\Omega_{2}\rho_{12}}{-i\Delta_{1}+\gamma}, \quad \rho_{23} =  \frac{\frac{i}{2}\Omega_{2}+i\Omega_{1}\rho_{21}}{-i\Delta_{2}+\gamma},
\end{equation} where the ground-state coherence is $\rho_{12}=\frac{-\Omega_{1}\Omega_{2}/\gamma}{\gamma_{12}+\frac{\Omega_{1}^2+\Omega_{2}^2}{\gamma}
	+i(\Delta_{2}-\Delta_{1})}$. The two optical susceptibilities are given by $ \chi_1 = \frac{N_0}{\varepsilon_0 E_R} ( d_{13} \rho_{13} + d_{14} \rho_{14}), \chi_2 = \frac{N_0}{\varepsilon_0 E_L} ( d_{23} \rho_{23} - d_{24} \rho_{24})$, and the real part $\mathrm{Re}(\chi)$ determines the refractive index. The phase difference between the left- and right-circularly polarized components after propagation through the atomic medium of length $L$ is $\phi = \frac{2\pi L(n_{R} - n_{L})}{\lambda}$, where $\lambda$ is the laser wavelength. The polarization state of the laser field is characterized by the Stokes vector, and the polarization rotation can be written as: 

\begin{equation}
	\label{eq:Sy}
	\begin{aligned}
		S_y &\propto [\mathrm{Re}(\chi_{1}) - \mathrm{Re}(\chi_{2})] \\
		&\propto \frac{1}{\gamma^2+\Delta_1^{2}}[( \Delta_{2} -  \Delta_{1}) (\frac{1}{2} +\mathrm{Re}(\rho_{12}))-2\gamma\mathrm{Im}(\rho_{12})].
	\end{aligned}
\end{equation}

Substituting the expression for $\rho_{12}$ into Eq.~\eqref{eq:Sy} yields:

\begin{equation}
	\begin{aligned}
		S_{y} &\propto (\Delta_{2}-\Delta_{1})(1-\frac{1}{2}\frac{\Omega^{2}}{(\gamma_{12}+\frac{\Omega^2}{\gamma})^{2}}) \\
		&\propto S_{z}\Omega^{2}(1-\frac{1}{2}\frac{\Omega^{2}}{(\gamma_{12}+\frac{\Omega^2}{\gamma})^{2}}),
	\end{aligned}
\end{equation} where $S_{z}= \frac{\Omega_{1}^{2}-\Omega_{2}^{2}}{\Omega_{1}^{2}+\Omega_{2}^{2}}$. This expression provides a formal derivation of the one-axis-twisting (OAT) Hamiltonian, $H\propto S_{z}^2$, which generates the shearing dynamics and squeezing of the quantum noise. 

\section*{Supplementary Note 2: Numerical simulation}

\subsection*{Multi-channel numerical model}

\begin{figure*}
	\centering
	\includegraphics[width=0.5\textwidth]{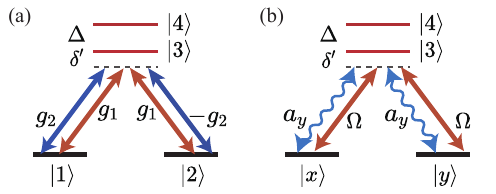}
	\caption{The double-$\Lambda$ theoretical model. (a) In the circular basis, the states $|1\rangle$ and $|2\rangle$ represent the Zeeman sublevels of the ground state $|5S_{1/2},F=2\rangle$. $g_1$ is the coupling strength for the transition $|5S_{1/2},F=2\rangle \rightarrow |5P_{1/2},F'=1\rangle$, and $g_2$ for the transition between $|5S_{1/2},F=2\rangle$ and $|5P_{1/2},F'=2\rangle$. (b) In the linear polarization basis, $\Omega$ is the Rabi frequency of the $x$-polarized pump field, and $a_y$ corresponds to the quantum field with $y$-polarization.}
	\label{Fig. S11}
\end{figure*}

We can quantitatively evaluate the quantum noise of $y$-polarized photons generated through the squeezing process by the input $x$-linearly polarized light. The Hamiltonian for the double-$\Lambda$ system  in the linear polarization basis (Fig. \ref{Fig. S11}(b)) can be expressed as:

\begin{equation}
	H_{int} = \frac{\hbar N_{0}}{L} \int_{0}^{L} dz \big[ \delta' \hat{\rho}_{33} + (\delta' + \Delta) \hat{\rho}_{44} - g_2 \alpha \hat{\rho}_{3y} - g_1 \alpha \hat{\rho}_{4x} - g_2 \hat{a}_y \hat{\rho}_{3x} - g_1 \hat{a}_y \hat{\rho}_{4y} + \mathrm{H.c.} \big],
\end{equation} 
where the linear polarization atomic bases are defined as $|x\rangle=(|1\rangle+|2\rangle)/\sqrt{2}$ and $|y\rangle=(|1\rangle-|2\rangle)/\sqrt{2}$. The single-photon Rabi frequency is given by $g_i = \mu_{i} \mathcal{E}/\hbar$, with $\mu_i$ denoting the dipole moment and $\mathcal{E}$ representing the single-photon electric field amplitude. The parameter $\alpha = E_x/\sqrt{\hbar\omega/(4\pi\epsilon_0 A)}$ characterizes the strength of the driving field, where $E_x$ is the amplitude of the $x$-polarized electric field, $\epsilon_0$ is the vacuum permittivity and $A$ denotes the cross-sectional area of the laser beam. $\delta'$ is the one-photon detuning and $\Delta$ is the hyperfine splitting between the two upper excited states. For an $N$-channel array where coherence in the dark region (not illuminated by the laser array) $\rho^{(0)}$ is treated as a non-Markovian bath, the evolution equations are:
\begin{equation}
	\begin{aligned}
		\begin{cases}
			\dot{\rho}^{(1)} &= -\frac{i}{\hbar}[H_{int}^{(1)}, \rho^{(1)}] - \frac{1}{2} \{\Gamma^{(1)}, \rho^{(1)}\} + S_1 - k_{10}\rho^{(1)} + k_{10}\rho^{(0)} + \mathfrak{F}_1 \\
			\dot{\rho}^{(2)} &= -\frac{i}{\hbar}[H_{int}^{(2)}, \rho^{(2)}] - \frac{1}{2} \{\Gamma^{(2)}, \rho^{(2)}\} + S_2 - k_{20}\rho^{(2)} + k_{20}\rho^{(0)} + \mathfrak{F}_2 \\
			&\vdots \\
			\dot{\rho}^{(N)} &= -\frac{i}{\hbar}[H_{int}^{(N)}, \rho^{(N)}] - \frac{1}{2} \{\Gamma^{(N)}, \rho^{(N)}\} + S_N - k_{N0}\rho^{(N)} + k_{N0}\rho^{(0)} + \mathfrak{F}_N \\
			\dot{\rho}^{(0)} &= -\frac{1}{2} \{\Gamma^{(0)}, \rho^{(0)}\} - \gamma_0\rho^{(0)} + S_0 - \sum_{i=1}^{N}k_{i0}\rho^{(0)} + \sum_{i=1}^{N}k_{0i}\rho^{(i)} + \mathfrak{F}_0
		\end{cases}
	\end{aligned}
\end{equation}
Here the relaxation matrix $\Gamma^{(i)}$ describes atomic decay processes, and $S_{i}$ represents the corresponding re-population dynamics. The forms of $S_{i}$ and $\Gamma^{(i)}$ are:

\begin{equation*}
	S_{i} = \begin{pmatrix}
		\frac{\gamma_0}{2} ( \rho_{yy}^{(i)} + \rho_{55}^{(i)} ) + \frac{38\gamma}{100} ( \rho_{33}^{(i)} + \rho_{44}^{(i)} ) & 0 & 0 & 0 & 0 \\
		0 & \frac{\gamma_0}{2} ( \rho_{xx}^{(i)} + \rho_{55}^{(i)} ) + \frac{38\gamma}{100} ( \rho_{33}^{(i)} + \rho_{44}^{(i)} ) & 0 & 0 & 0 \\
		0 & 0 & 0 & 0 & 0 \\
		0 & 0 & 0 & 0 & 0 \\
		0 & 0 & 0 & 0 & \frac{24\gamma}{100} ( \rho_{33}^{(i)} + \rho_{44}^{(i)} ) + \frac{\gamma_0}{2} ( \rho_{xx}^{(i)} + \rho_{yy}^{(i)} )
	\end{pmatrix},
\end{equation*}
\begin{equation}
	\Gamma^{(i)} = \begin{pmatrix}
		\gamma_0 & \gamma_{12} & \frac{\gamma}{2} & \frac{\gamma}{2} & \gamma_{12} \\
		\gamma_{12} & \gamma_0 & \frac{\gamma}{2} & \frac{\gamma}{2} & \gamma_{12} \\
		\frac{\gamma}{2} & \frac{\gamma}{2} & \gamma & \gamma & \frac{\gamma}{2} \\
		\frac{\gamma}{2} & \frac{\gamma}{2} & \gamma & \gamma & \frac{\gamma}{2} \\
		\gamma_{12} & \gamma_{12} & \frac{\gamma}{2} & \frac{\gamma}{2} & \gamma_0
	\end{pmatrix}.
\end{equation}

The hopping rates between an optical channel and the dark region are given by $k_{i0} = \frac{dN_{i}}{N_idt} = \frac{2\pi r_i}{\pi r_i^2} \left( \frac{m}{2\pi k_B T} \right)^{1/2} \int_0^{+\infty} e^{-\frac{mv_z^2}{2k_B T}} v_z \, dv_z \approx \frac{\bar{v}}{2r_i} \sim 10^5 s^{-1}$ and $k_{0i} = \frac{dN_i}{N_ddt} = k_{i0} \frac{r_i^2}{R_{cell}^2-\sum_i{r_i^2}} \sim 10^3 s^{-1}$ respectively, where $r_i$ is the radius of the i-th channel, $R_{cell}$ is the radius of the vapor cell, $k_B$ is the Boltzmann constant, $v$ is the velocity of atoms, $m$ is the mass of a rubidium atom, $T$ is the temperature, $N_i$ is the atom number in the $i$-th channel and $N_d$ is the atom number in the dark region. The Langevin operator $\mathfrak{F}_{(i)}$ characterizes the environment induced fluctuations of the atomic operators, which directly drives the atomic noise dynamics and affects the quantum properties of the output light field. In addition, the evolution of the vacuum-field is governed by
\begin{equation}
	\left( \frac{\partial}{\partial t} + c\,\frac{\partial}{\partial z} \right) 
	\hat{a}_{y}(z, t) = \frac{1}{i\hbar} \left[ \hat{a}_{y}(z, t), H_{int} \right].
\end{equation}
For a multi-channel configuration, the propagation equations for the corresponding vacuum-field take the form:
\begin{equation}
	\begin{cases}
		\begin{aligned}
			\left( \frac{\partial}{\partial t} + c\,\frac{\partial}{\partial z} \right) 
			\hat{a}_{y}^{(1)}(z, t) &= i g_{1} N_{0}\, \hat{\rho}_{y4}^{(1)} 
			+ i g_{2} N_{0}\, \hat{\rho}_{x3}^{(1)} \\
			\left( \frac{\partial}{\partial t} + c\,\frac{\partial}{\partial z} \right) 
			\hat{a}_{y}^{(2)}(z, t) &= i g_{1} N_{0}\, \hat{\rho}_{y4}^{(2)} 
			+ i g_{2} N_{0}\, \hat{\rho}_{x3}^{(2)}\\
			&\vdots \\
			\left( \frac{\partial}{\partial t} + c\,\frac{\partial}{\partial z} \right) 
			\hat{a}_{y}^{(N)}(z, t) &= i g_{1} N_{0}\, \hat{\rho}_{y4}^{(N)} 
			+ i g_{2} N_{0}\, \hat{\rho}_{x3}^{(N)}
		\end{aligned}
	\end{cases}
\end{equation}
Once the evolution equations for both the atomic and optical field operators are established, the system dynamics can be investigated in detail. To study the quantum fluctuations of the light field, we then linearize the dynamics around the steady state by expanding the operators as $\hat{\rho}_{\mu\nu}^i = \langle \hat{\rho}_{\mu\nu}^i \rangle + \delta\hat{\rho}_{\mu\nu}^i$ and $\hat{a}_y = \langle \hat{a}_y \rangle + \delta\hat{a}_y$, where $\langle \hat{a}_y \rangle=0$. Substituting these into the evolution equations and transforming to the Fourier domain, we solve for the fluctuation $\delta\hat{a}_y(\omega)$ in terms of the atomic operators. The numerical solution yields the spectrum of the output vacuum field, which contains the quantum noise contributions from both the atomic and field fluctuations. The squeezing is quantified by the noise spectrum of a general quadrature, $\hat{X}_\theta = \hat{X}\cos\theta + \hat{P}\sin\theta$, where $\hat{X} = (\hat{a}_y + \hat{a}_y^\dagger)/2$ and $\hat{P} = (\hat{a}_y - \hat{a}_y^\dagger)/(2i)$ are the amplitude and phase quadratures, respectively. The noise spectrum of a given quadrature $\hat{X}_{\theta}$ can be obtained from the Fourier transform of its auto-correlation function $S_{X_{\theta}}(\omega) = \left\langle \hat{X}_{\theta}(\omega)\, \hat{X}_{\theta}(\omega') \right\rangle$. The minimum value of $S_{X_{\theta}}(\omega)$ at the angle $\theta$ identifies the squeezing level at frequency $\omega$, and the corresponding $\theta_{sq}$ defines the squeezing angle.

To accurately model the atomic dynamics and squeezing generation in the rubidium vapor cell, we extend the standard double-$\Lambda$ model to include an auxiliary energy level that accounts for population loss induced by optical pumping. This is necessary since the actual ${}^{87}$Rb atoms feature an additional hyperfine level $|F=1\rangle$ that remains uncoupled with the applied optical field.
For a four-level model, the Hamiltonian is:

\begin{equation}
	H_{int} = -\hbar
	\begin{pmatrix}
		0 & 0 & \hat{a}_{y}\,g_1 & \sqrt{2}\,\alpha\,g_2 \\
		0 & 0 & \sqrt{2}\,\alpha\,g_1 & \hat{a}_{y}\,g_2 \\
		\hat{a}^{\dagger}_{y}\,g_1 & \sqrt{2}\,\alpha\,g_1 & -\delta & 0 \\
		\sqrt{2}\,\alpha\,g_2 & \hat{a}^{\dagger}_{y}\,g_2 & 0 & -(\delta+\Delta)
	\end{pmatrix};
\end{equation} 
while for a five-level model including the auxiliary state with a hyperfine splitting of $\Delta_{HF} = 6.8$ GHz, the Hamiltonian becomes:
\begin{equation}
	H_{int} = -\hbar
	\begin{pmatrix}
		0 & 0 & \hat{a}_{y}\,g_1 & \sqrt{2}\,\alpha\,g_2 & 0 \\
		0 & 0 & \sqrt{2}\,\alpha\,g_1 & \hat{a}_{y}\,g_2 & 0 \\
		\hat{a}^{\dagger}_{y}\,g_1 & \sqrt{2}\,\alpha\,g_1 & -\delta & 0 & 0 \\
		\sqrt{2}\,\alpha\,g_2 & \hat{a}^{\dagger}_{y}\,g_2 & 0 & -(\delta+\Delta) & 0 \\
		0 & 0 & 0 & 0 & 2\pi \times \Delta_{HF}
	\end{pmatrix};
\end{equation}

In our numerical simulations, we use a Rb cylindrical vapor cell with a diameter of 2.5 cm and length of 7.5 cm, a decay rate of $\gamma_0 = \gamma_{12} =10$ Hz, and a Doppler width of $\Delta_D = 500$ MHz, unless otherwise stated. The model describes the intrinsic squeezed-state generation from the nonlinear atom-light interaction and does not include external optical loss or technical noise. It therefore reproduces the main physical trends only qualitatively, while some deviation in the absolute values from the experimental data is expected.

\subsection*{Effects of Doppler broadening on squeezing angle in four-level model}

\begin{figure*}
	\centering
	\includegraphics[width=0.7\textwidth]{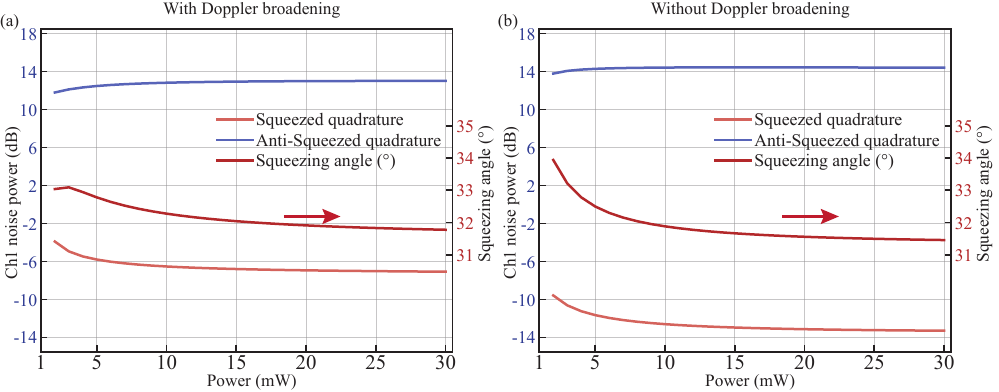}
	\caption{Four-level model numerical results for the single-channel case, showing the squeezed and anti-squeezed quadrature noise powers and the squeezing angle as a function of laser power. (a) Four-level atomic model with Doppler broadening included. (b) Four-level atomic model without Doppler broadening. The cell temperature is set at $63.7$$^{\circ}$C, and the laser frequency is set to be $80$ MHz red-detuned from $|5S_{1/2}F=2\rangle \rightarrow |5P_{1/2}F'=1\rangle$. Cell length: 7.5 cm; diameter: 2.5 cm. A laser power of $1$ mW corresponds to a Rabi frequency of $43.6$ MHz under the theoretically chosen parameters. The detection frequency is fixed at $160$ kHz, and the noise power is normalized to the shot noise level ($0$ dB level).}
	\label{Fig. S12}
\end{figure*}

Both theoretical and experimental results confirm the presence of substantial excess noise above the shot-noise limit, especially in the anti-squeezed quadrature, contrasting with the balanced degree of squeezing and anti-squeezing a unitary squeezed state. This excess noise arises primarily from dissipative processes, particularly spontaneous emission of the atoms associated with the absorption of light. When the excess noise dominates, the small variation of the squeezing angle with experimental parameters cannot be detected. To verify this, we calculate the quantum noise and the squeezing angle with and without Doppler broadening of the excited states, within the four-level model and the single-channel case. Doppler broadening is included by integrating the system’s response over the Maxwell-Boltzmann velocity distribution of the atoms. In presence of Doppler broadening, EIT efficiency is lower and the spontaneous emission introduces more excess noise. As illustrated in Fig. \ref{Fig. S12}(a) and (b), one can observe that as the laser power changes, the squeezing angle varies accordingly. As expected, however, the variation is more pronounced without Doppler broadening, where the squeezed state is also closer to a unitary one, exhibiting lower excess noise in the anti-squeezed quadrature.

\subsection*{Comparison of four-level and five-level models}

\begin{figure*}
	\centering
	\includegraphics[width=0.7\textwidth]{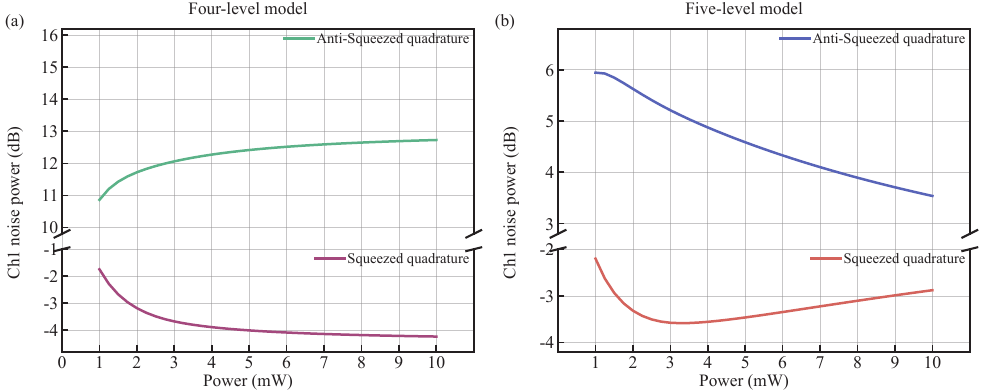}
	\caption{Numerical results of squeezed and anti-squeezed quadratures in the single-channel case as a function of the laser power. (a) The four-level model results. (b) The five-level model results (three ground states and two excited states, while an auxiliary ground state acts as the population trapping state). The cell temperature is set at $63.7$$^{\circ}$C, and the laser is on resonant with $|5S_{1/2}F=2\rangle \rightarrow |5P_{1/2}F'=1\rangle$. A laser power of $1$ mW corresponds to a Rabi frequency of $43.6$ MHz under the theoretically chosen parameters. Cell length: 7.5 cm; diameter: 2.5 cm. The detection frequency is fixed at $160$ kHz, and the noise power is normalized to the shot noise level ($0$ dB level).}
	\label{Fig. S9}
\end{figure*}

To further demonstrate that light absorption is the source of excess noise, we compare the four-level and five-level models. While the four-level model accurately captures the mechanism of squeezing generation in the atomic system, the actual atomic ground state includes both $|F=1\rangle$ and $|F=2\rangle$ hyperfine levels, indicating that the five-level model is more realistic. Due to the auxiliary state in the five-level model that can trap the leaked population via optical pumping, light absorption for high laser power is reduced compared to that in the four-level model under the same parameters. For comparison between the two models, the calculated laser power dependence of the quantum noise for the single channel case is presented in Fig. \ref{Fig. S9}(a) and (b). In our calculations, identical parameters are employed in the two models for direct comparison. The temperature is chosen to match that used in the experiment. The decay rates of ground-state population and coherence are set to $10$ Hz, where we assume that the ground-state coherence decay occurs in the dark region due to wall collisions and small magnetic field inhomogeneity. Our calculations account for Doppler broadening and use an excited-state decay rate of $6$ MHz. The four-level model results (Fig. \ref{Fig. S9}(a)) demonstrate the characteristic PSR enhancement with increasing power, accompanied by simultaneous increase of the degree of squeezing and anti-squeezing, where their asymmetric deviations from the photon shot noise level are due to the excess atomic noise. In contrast, the five-level model in Fig. \ref{Fig. S9}(b) exhibits a distinct behavior: the anti-squeezing noise decreases with increasing optical power, in agreement with our experimental observations. The drop in the quantum noise in the anti-squeezed quadrature can be attributed to more pronounced population leakage from the $|F=2\rangle$ state at higher laser powers, which reduces the effective atom number participating in the light-atom interactions, consequently suppressing the atomic noise contribution to the anti-squeezed component.

\subsection*{Performance of the squeezed array}

\begin{figure*}
	\centering
	\includegraphics[width=0.5\textwidth]{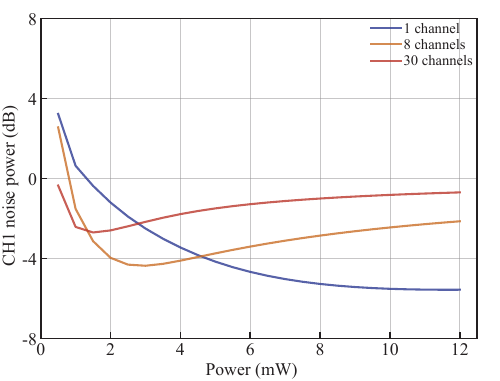}
	\caption{Numerical results of the squeezed quadrature noise power of CH1 as a function of laser power in one-, eight-, and thirty-channel arrays, where each channel in the array has identical input laser power. The temperature is $63.7$$^{\circ}$C, with the laser frequency set to be on resonance with $|5S_{1/2}F=2\rangle \rightarrow |5P_{1/2}F'=1\rangle$. Cell length: 7.5 cm; diameter: 2.5 cm. A laser power of $1$ mW corresponds to a Rabi frequency of $44.2$ MHz under the theoretically chosen parameters. The detection frequency is fixed at $160$ kHz, and the noise power is normalized to the shot noise level ($0$ dB level).}
	\label{Fig. S1}
\end{figure*}

Based on the five-level model, we perform numerical simulations for the array while keeping all other parameters identical to those described above for the single-channel case in Fig. \ref{Fig. S9}. Fig. \ref{Fig. S1} shows the squeezing versus laser power for single-channel, $8$-channel, and $30$-channel configurations, which agree well with the experimental results in Fig. 2(a) of the main text. For equal incident laser power per channel, the required power for a specific level of squeezing decreases with increasing array size, while larger arrays exhibit earlier saturation due to higher total laser power. 

\begin{figure*}
	\centering
	\includegraphics[width=0.5\textwidth]{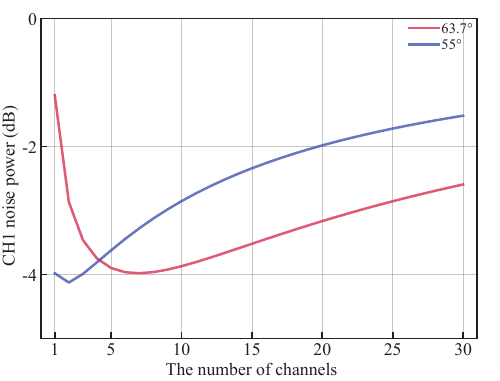}
	\caption{Numerical results of the squeezed quadrature noise power of CH1 as a function of the array size at temperatures of $63.7$$^\circ$C and $55$$^\circ$C, where each channel in the array has the same input power of $2$ mW. The laser frequency is set to be on resonant with $|5S_{1/2}F=2\rangle \rightarrow |5P_{1/2}F'=1\rangle$. A laser power of $2$ mW corresponds to a Rabi frequency of $62.6$ MHz under the theoretically chosen parameters. Cell length: 7.5 cm; diameter: 2.5 cm. The detection frequency is fixed at $160$ kHz, and the noise power is normalized to the shot noise level ($0$ dB level).}
	\label{Fig. S2}
\end{figure*}

The numerical results shown in Fig. \ref{Fig. S1} reveal that the maximum achievable squeezing in large-scale arrays is lower, compared to the single-channel configuration. This is limited by the total number of atoms available, as experimentally verified by the results presented in the inset of Fig. 2(b) in the main text. As shown in Fig. \ref{Fig. S2}, our theoretical results also illustrate the dependence of quantum noise on the array size at $63.7$$^\circ$C and $55$$^\circ$C, in agreement with the experimental results in the main text. At higher temperatures, saturation with array size is delayed as a result of the increase in total atom number, and the squeezing performance for a large array improves with temperature. This suggests that increasing temperature is a practical route for optimizing squeezing in array.

\subsection*{Squeezing angle}

\begin{figure*}
	\centering
	\includegraphics[width=0.5\textwidth]{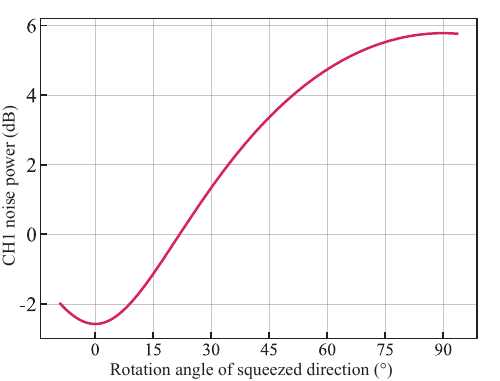}
	\caption{Calculated noise power of the quadrature for a combined beam of two identical squeezed states (with squeezing level of $2.58$ dB and anti-squeezing level of $5.76$ dB). The squeezing angle of one squeezed state is fixed, while the squeezing angle of the other is varied.}
	\label{Fig. S3}
\end{figure*}

In the main text, we described the experiments of measuring the squeezing of two combined (at the cell output) channels with distinct squeezing angles. Theoretically, we calculate the total quantum noise for two squeezed states with different phases in the squeezed vacuum. The results are presented in Fig. \ref{Fig. S3}, which are in good agreement with Fig. 5(b) of the main text. The degree of squeezing reaches maximum when the relative phase of two squeezed vacuum state is $0^\circ$, and reaches minimum when the relative phase is $90^\circ$. When the squeezing direction of the two states are aligned, the total variance $\mathrm{Var}(X)=\mathrm{Var}(X_{1})=\mathrm{Var}(X_{2})$. In contrast, when the squeezing directions are orthogonal (separated by $n\pi/2$), the variance $\mathrm{Var}(X)=(\mathrm{Var}(X_{1})+\mathrm{Var}(P_{2}))/2$.

\section*{Supplementary Note 3: Additional experimental results}

\subsection*{Balanced homodyne detection and shot-noise calibration}

\begin{figure*}
	\centering
	\includegraphics[width=0.7\textwidth]{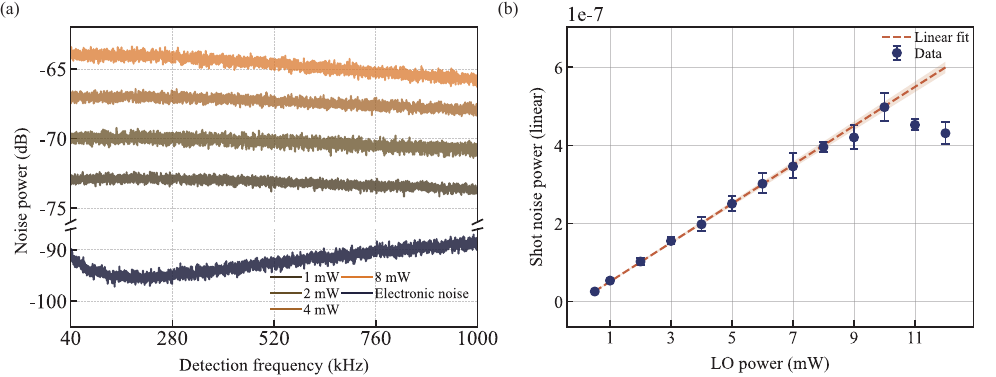}
	\caption{Noise characterization of balanced homodyne detection. (a) Representative raw noise spectra measured with the balanced homodyne detector, including the electronic noise floor and shot-noise spectra at several LO powers. (b) Shot-noise power versus LO power. The data points are obtained by averaging the spectrum around the analysis frequency 180 kHz, and the dashed line is a linear fit in linear-power units. Error bars indicate the standard deviation within the averaging window, and the shaded region around the dashed line shows the fit uncertainty.}
	\label{Fig. S13} 
\end{figure*}

Quantum quadrature noise is measured with a balanced homodyne detector (Thorlabs PDB450A; transimpedance gain, $10^5$ V/A), using the $x$-polarized pump field as the local oscillator (LO). Each squeezing measurement is preceded by an independent shot-noise calibration at the corresponding LO power.
A polarizing beam splitter (PBS) separated the $y$-polarized squeezed vacuum field (reflected port) from the $x$-polarized LO (transmitted port). During shot-noise calibration, coherent vacuum field enters the second PBS input port, and the recorded noise is taken as the shot-noise reference. By varying the relative phase between the squeezed vacuum field and the LO, the quantum noise of arbitrary quadratures is measured, while ensuring that the LO power is equal to that used for the corresponding shot-noise calibration. The noise power spectral density is recorded as a function of LO power from $0.5$ to $12$ mW with fixed spectrum-analyser settings. To quantify the shot-noise scaling, the noise power is extracted by averaging the spectrum over a narrow band centred at $180$ kHz, chosen as a representative analysis frequency. A linear fit of the noise power (in linear units) versus LO power confirms shot-noise-limited operation. The electronic noise floor remains significantly below the shot-noise over the full data acquisition bandwidth, indicating a negligible contribution to the reported squeezing data. For an LO power of $3.188$ mW, the detector imbalance is $0.31\%$. Representative raw noise spectra at several LO powers, together with the electronic noise floor, are shown in \ref{Fig. S13}(a), and the corresponding noise variance versus LO power, the linear fit and its uncertainty are shown in \ref{Fig. S13}(b).

\subsection*{Influence of the laser beam size}

\begin{figure*}
	\centering
	\includegraphics[width=0.7\textwidth]{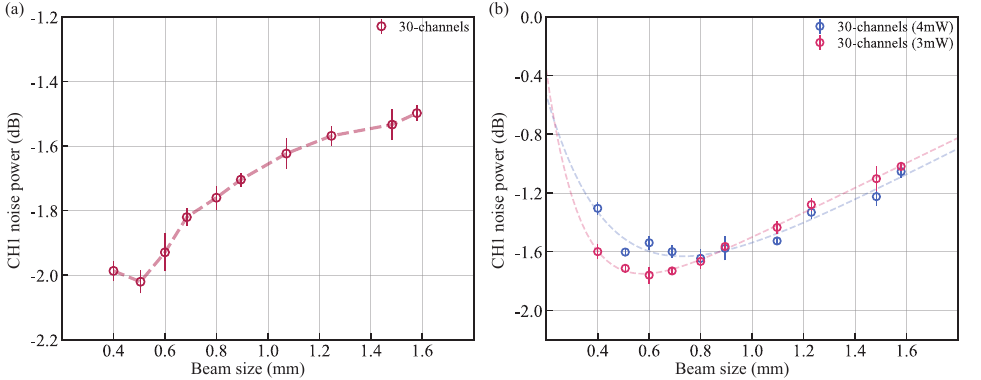}
	\caption{Experimental light squeezing in a 30-channel array versus the laser beam size. (a) The optimal squeezing for various beam sizes, $0.400$ mm, $0.500$ mm, $0.600$ mm, $0.689$ mm, $0.800$ mm, $0.895$ mm, $1.097$ mm, $1.230$ mm, $1.483$ mm, and $1.579$ mm, with the corresponding optimal laser power of $1.75$ mW, $2$ mW, $2.25$ mW, $2.5$ mW, $2.75$ mW, $3.2$ mW, $3.5$ mW, $3.5$ mW, $4.0$ mW, and $3.5$ mW per channel, respectively. The experiment is conducted at a cell temperature of $63.7$$^\circ$C. (b) The squeezing versus  beam size for fixed input powers of $3$ mW and $4$ mW (per channel) respectively, using the same set of beam sizes as in (a). The detection frequency is $160$ kHz, and error bars represent the standard deviation from more than six repeated measurements. The noise power is normalized to the shot noise level ($0$ dB).}
	\label{Fig. S4}
\end{figure*}

In addition to the parameters discussed in Fig. 2 of the main text, laser beam size is another critical factor determining the squeezing performance of the array. As shown in Fig. \ref{Fig. S4}(a), the maximum achievable squeezing varies with the beam size, with an optimal value of about $0.5$ mm for the 30-channel array. This is due to a trade off between competing effects. A larger beam size strengthens the inter-channel coupling but reduces the optical power density for a fixed total power. Meanwhile, a smaller beam increases the power density but eventually leads to saturation, which degrades the squeezing. 

The interplay between the beam size and the laser power is further detailed in Fig. \ref{Fig. S4}(b), which plots the noise power of a chosen channel against the array's beam size for two different input powers ($3$ mW and $4$ mW per channel) in a 30-channel array. For beam sizes below $0.8$ mm, a lower power of $3$ mW yields superior squeezing. In this regime, the higher total power at $4$ mW induces more spontaneous emission noise that degrades the squeezing. Conversely, for beams larger than $1.2$ mm, the higher power of $4$ mW becomes advantageous. Here, the system is limited by the efficiency of the four-wave-mixing process. The higher $4$ mW power is required to achieve sufficient nonlinear interaction because of the lower optical density at larger beam sizes.  This identifies the overall optimal squeezing performance for the array corresponds to the regime of relatively lower total power and smaller beam size.

\subsection*{Investigation of the anti-squeezed quadrature}

\begin{figure*}
	\centering
	\includegraphics[width=0.7\textwidth]{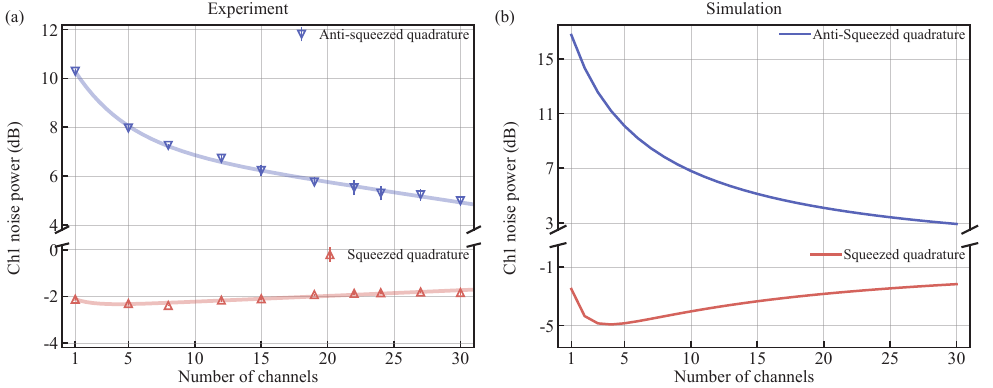}
	\caption{Experimental and numerical noise power of the squeezed and anti-squeezed quadratures versus the array's channel numbers, for laser power of $3$ mW per channel. The beam size is $0.505$ mm, and the cell temperature is $63.7$$^\circ$C. The detection frequency is $180$ kHz, and the noise power is relative to the shot noise level ($0$ dB). Error bars are obtained from more than six repeated measurements. The numerical results are obtained under the same parameters as specified above, employing a five-level model with a Rabi frequency of $76.6$ MHz.}
	\label{Fig. S5}
\end{figure*}

In Fig. 4 of the main text, we show the squeezed and anti-squeezed quadrature noises versus the array's channel numbers, at relatively lower laser power ($1$ mW per channel). Here Fig. \ref{Fig. S5} extends these results to higher input light power ($3$ mW per channel). At $3$ mW, the noise power in the squeezed quadrature decreases first and then rises as the number of channels in the array increases, while the noise power in the anti-squeezed quadrature consistently decreases. The numerical calculations are in good qualitative agreement with the experimental trends. 

\subsection*{Optical phase distribution of the array}

\begin{figure*}
	\centering
	\includegraphics[width=0.8\textwidth]{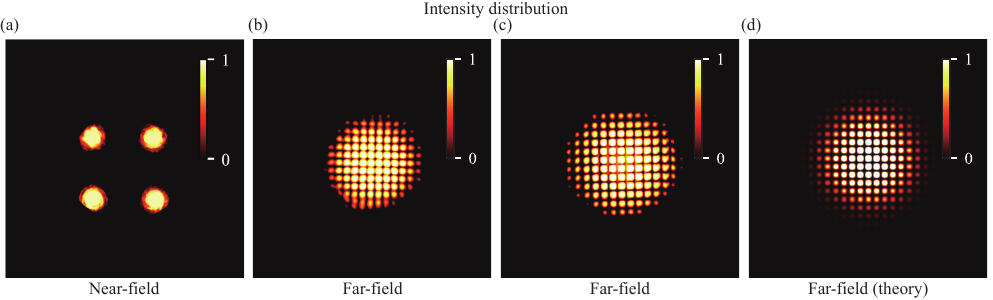}
	\caption{Near-field and far-field images of 4-beam lattice in the 30-channel array. 
		(a) Measured near-field intensity distribution of four adjacent beams in the array, showing a square lattice configuration with uniform beam size and intensity. Each channel is represented by a dot captured by a camera. 
		(b) Measured far-field intensity distribution of a 4-beam lattice before entering the rubidium vapor cell. 
		(c) Measured far-field intensity distribution of a 4-beam lattice after exiting the rubidium vapor cell.
		(d) Theoretically calculated far-field intensity distribution of a in-phase 4-beam lattice.
	}
	\label{Fig. S7}
\end{figure*}

In the main text, we mentioned that the relative phase between the $x$-polarized field component and the squeezed vacuum determines the squeezing angle of the ellipse on the Poincar\'{e} sphere for the polarization squeezed state. Therefore, if the squeezing angles of all channels in the array are the same, and if the input $x$-polarized (pump) array are in phase, then we could infer that the squeezed vacuum array are in phase too. In order to prove that the pump array are in phase, we take images of both the near-field and far-field planes by a CCD camera. The far-field plane, representing the Fourier transform of the near-field, can reveal the information of the phase distribution of the array. When the optical channels are in phase, interference in the far field produces a well-defined intensity pattern~\cite{mahler2020improved}; conversely, random phases result in a disordered distribution. Due to the finite size of the lens and the relatively large spacing between the optical channels in the array, spherical aberration affects the imaging quality. Therefore, we randomly select (by an aperture) four adjacent optical channels arranged in a $2\times2$ square lattice out of the 30-channel array for measurement. The near-field intensity distribution is shown in Fig. \ref{Fig. S7}(a), indicating uniform power across all channels. While the far-field intensity distribution (before the beams enter the rubidium vapor cell) is shown in Fig. \ref{Fig. S7}(b), demonstrating that the optical array maintains near-perfect identical phases. After propagation through the rubidium vapor cell, we again arbitrarily select four channels for imaging, with the results presented in Fig. \ref{Fig. S7}(c), confirming that the array retains its phase stability. Fig. \ref{Fig. S7}(d) shows the theoretically calculated far-field intensity distribution of a 4-beam lattice where all the light beams have the same phase. It resembles the experimentally measured far-field intensity distribution pattern, indicating that the pump array are in phase.  

\section*{Supplementary Note 4: Routes to improved performance of the squeezed array}

In this section we point out, mainly through numerical simulations, the main factors affecting the squeezing level for the array and routes for future improvements. Overall, the guiding principle for larger squeezing in the array is to increase the total number of atoms, by increasing the atomic density (higher vapor cell temperature), using a longer or wider vapor cell. Then under a particular set of conditions of cell geometry and temperature, the laser power and detuning should be optimized accordingly.

\subsection*{Temperature dependence and wall coating}

The maximal achievable squeezing level depends on the strength of the nonlinear atom-light interaction, which connects to the atomic vapor density and hence the cell temperature. Increasing the temperature enhances the optical depth (OD) and can improve squeezing. This dependence is evident in the experimental data in the inset of Fig.~2(b): raising the temperature from $50$$^\circ$C to $63.7$$^\circ$C yields $\sim$1 dB increase in squeezing, with the optimal pump laser power per beam changing from 0.75 mW to 2 mW. We emphasize that the temperature ($63.7$$^\circ$C) optimally chosen in our experiment is limited by the paraffin coating on the cell's inner wall, and can be further increased if the coating material could tolerate higher temperatures. When the temperature approaches the melting point of the coating (typically 60–70$^\circ$C for the paraffin in our cell), its capability of preserving the atomic coherence during wall collisions degrades~\cite{Li2017JAP}, leading to higher spin relaxation rates. In addition, temperature-dependent modifications of atom-surface interactions, including adsorption and desorption dynamics, can reduce the effective vapor density~\cite{Atutov2017EPJD}. Consequently, these effects impose a practical upper limit for the operating temperature of the squeezed array.

\begin{figure*}
	\centering
	\includegraphics[width=0.5\textwidth]{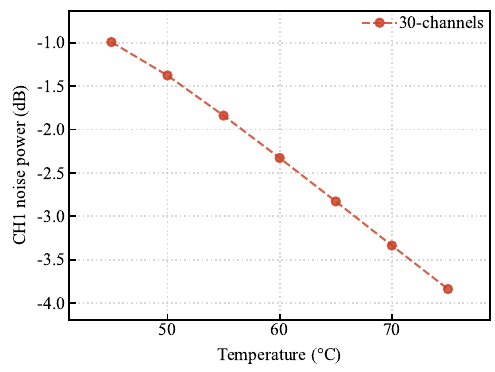}
	\caption{Numerical results for the squeezed quadrature noise power of CH1 in a 30-channel array as a function of temperature where each channel in the array has the same input power. The laser frequency is resonant with $|5S_{1/2}F=2\rangle \rightarrow |5P_{1/2}F'=1\rangle$. The detection frequency is fixed at $160$ kHz, and the noise power is normalized to the photon shot noise level ($0$ dB level). The cell geometry is fixed (length: 7.5 cm, diameter: 2.5 cm). For each temperature, the optical power per beam is chosen close to its corresponding optimal value to produce the best achievable squeezing under that condition. The powers used are 0.75 mW, 0.75 mW, 1 mW, 1.25 mW, 1.75 mW, 2.25 mW, and 2.75 mW for temperatures of 45$^\circ$C, 50$^\circ$C, 55$^\circ$C, 60$^\circ$C, 65$^\circ$C, 70$^\circ$C, and 75$^\circ$C, respectively. The results show that higher temperature leads to higher optimal squeezing while the corresponding optimal laser power is higher, consistent with delayed onset of saturation for more atoms.}
	\label{Fig. S16}
\end{figure*}

We numerically calculate how the array squeezing can improve with increasing temperature, assuming that the coating property does not degrade, and the temperature is not too high to cause decoherence due to atom-atom collisions~\cite{optical-pumping}. As shown in Fig.~\ref{Fig. S16}, our model reproduces the experimentally observed trend of squeezing improvement with higher temperature (below $63.7$$^\circ$C). The simulation results suggest that squeezing can continue to improve between $63.7$$^\circ$C and $75$$^\circ$C. We note that, for larger OD, the onset of saturation shifts to higher pump laser powers, which results in a larger optimal operating laser power. For example, our simulation shows that as the temperature increases from 50$^\circ$C to 75$^\circ$C, the optimal power changes from $\sim$0.75 mW to $\sim$2.75 mW, while the corresponding squeezing level becomes higher.

This temperature-dependent saturation behavior occurs simply because more photons (higher total laser power in the cell) are needed to saturate more atoms (higher cell temperatures). It also accounts for the experimentally observed trends of array squeezing versus the number of channels shown in Fig.~2(b). At $55$$^\circ$C the squeezing degrades monotonically with increasing number of channels (under fixed per-channel power), indicating early onset of saturation. Whereas at $63.7$$^\circ$C the squeezing initially improves with the number of channels before reaching a plateau, suggesting that more atoms for the high temperature case delays the saturation.

In addition, we note that previous PSR squeezing experiments in Rb vapor cell (without wall coating) were conducted in the 70–85$^\circ$C range~\cite{Zhang2017PRA}, which qualitatively agree with our numerical results favoring higher temperatures. Accessing this higher temperature regime in our platform requires replacing the paraffin coating with a thermally more stable alternative such as octadecyltrichlorosilane (OTS) that can withstand temperatures up to $\sim$170$^\circ$C~\cite{seltzer2009high}. However, for PSR that involves near-resonant interaction and ground state coherence, a too-high temperature will cease to benefit squeezing because absorption and spin-exchange induced decoherence are severe~\cite{optical-pumping}.

\subsection*{Effect of the vapor cell length}

\begin{figure*}
	\centering
	\includegraphics[width=0.5\textwidth]{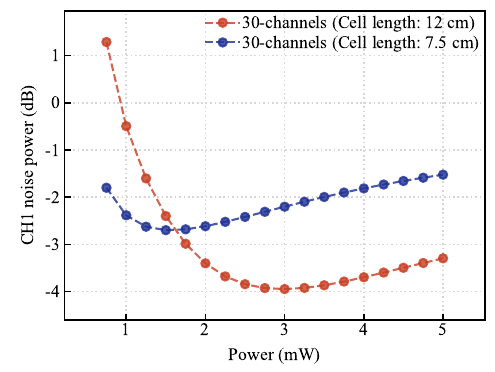}
	\caption{Squeezing at different cell lengths. Numerical results for the squeezed quadrature noise power of CH1 in a 30-channel array as a function of optical power per channel, for cell lengths of 7.5\,cm (blue) and 12\,cm (red). The laser is resonant with $|5S_{1/2}F=2\rangle \rightarrow |5P_{1/2}F'=1\rangle$. The detection frequency is fixed at $160$ kHz, and the noise power is normalized to the photon shot noise level ($0$ dB level). The cell diameter is fixed at 2.5 cm, and temperature is 63.7$^\circ$C. The results show that increasing the cell length shifts the optimal power to higher values and enables a higher maximum squeezing.}
	\label{Fig. S17}
\end{figure*}

Extending the vapor cell length increases the OD and strengthens the nonlinear atom-light interaction. Meanwhile, due to more atoms in a longer cell, more photons (higher total laser power in the cell) are required to saturate the atoms. For instance, as shown in Figure.~\ref{Fig. S17}, at 63.7$^\circ$C, our numerical simulation indicates that increasing the cell length from 7.5 cm to 12 cm (with cell diameter fixed at 2.5 cm) improves the maximum squeezing from $\sim$2.7 dB (at 1.5 mW/channel) to $\sim$3.95 dB (at 3 mW/channel).  We note that there is a limit for the cell length because excess noise due to spontaneous emission can build up during light propagation along the cell and eventually harm the squeezing. Therefore, the optimal cell length also depends on the laser power, cell temperature etc.

\subsection*{Influence of the width of the vapor cell}

In paraffin-coated vapor cells, atomic diffusion constantly redistributes ground-state population and coherence across the entire cell volume, enabling all atoms in the cell to collectively participate in the nonlinear atom-light interaction. As atoms in the optical channels are pumped into the trapped states (other hyperfine levels) and cease to participate the PSR process, fresh atoms can diffuse into the channels from the surrounding unilluminated region, which help to sustain the nonlinear optical process. Consequently, the saturation threshold is affected by the total atom number in the cell rather than the local atoms within any individual channel.

\begin{figure*}
	\centering
	\includegraphics[width=0.5\textwidth]{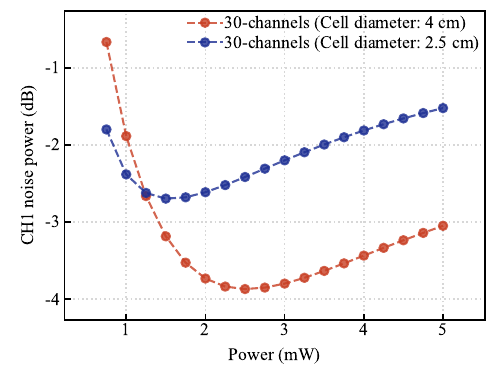}
	\caption{The influence of cell width on array squeezing. Numerical results for the squeezed quadrature noise power of CH1 in a 30-channel array as a function of optical power per channel, for cell diameters of 2.5 cm (blue) and 4 cm (red). The laser is resonant with $|5S_{1/2}, F{=}2\rangle \rightarrow |5P_{1/2}, F'{=}1\rangle$. Detection frequency: 160 kHz; noise power normalized to the photon shot noise level (0 dB). Cell length: 7.5 cm; temperature: 63.7$^\circ$C. Increasing cell diameter raises the total atom number, shifts the optimal pump power to higher values, and yields larger peak squeezing.}
	\label{Fig. S18}
\end{figure*}

For a fixed array size, increasing the cell width or the diameter creates a larger bath for fresh atoms which can better replenish atoms within the optical channels. This would raise the saturation threshold in terms of the per-channel laser power.  Figure.~\ref{Fig. S18} presents the squeezing versus laser power for two different cell diameters, showing that both the optimal power and peak squeezing increase with the cell diameter: at 63.7$^\circ$C, increasing the cell diameter from 2.5 cm to 4 cm (cell length fixed at 7.5 cm) can improve the maximal squeezing from $\sim$2.7 dB (reached at 1.5 mW/channel) to $\sim$3.87 dB (reached at 2.5 mW/channel). This confirms that the total atom number in the cell influences the onset point of the saturation behavior. Finally, we note that while wider cell is preferred for the squeezed light array, in practice, a wider vapor cell requires more stringent control of the residual magnetic field inhomogeneity within the shielding.

\subsection*{Multi-parameter optimization}

The squeezing performance reported in the main text was observed in the coated vapor cell available in the lab, and does not represent the ultimate limit of our scheme. As described above, the cell temperature and cell geometry such as the length and width, provide room for improvement. In addition, laser parameters such as the laser power and laser detuning should be optimized accordingly because the saturation threshold depends on the total atom number. Overall, it is a problem of multi-parameter optimization.

\begin{figure*}
	\centering
	\includegraphics[width=0.7\textwidth]{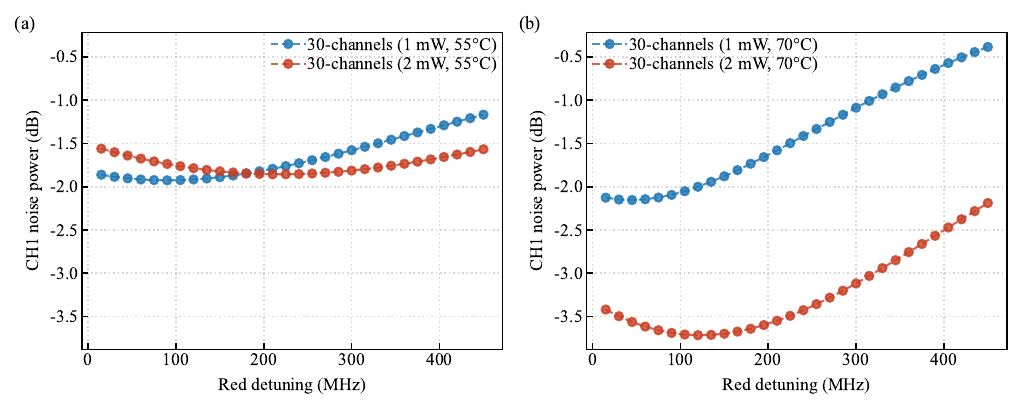}
	\caption{Effect of laser detuning on light squeezing. Numerically calculated squeezed-quadrature noise power for CH1 in a 30-channel array as a function of the value of the red detuning of the laser, for per-channel optical powers of 1 mW (blue) and 2 mW (red) at two temperatures. Zero detuning corresponds to resonance with the $|5S_{1/2}, F{=}2\rangle \rightarrow |5P_{1/2}, F'{=}1\rangle$ transition. Detection frequency: 160 kHz; Cell length: 7.5 cm; Cell diameter: 2.5 cm; Noise power is normalized to the photon shot noise level (0 dB). (a) Temperature: 55$^\circ$C. (b) Temperature: 70$^\circ$C. It can be seen that the optimal laser detuning has a dependence on laser power and temperature.}
	\label{Fig. S19}
\end{figure*}

The PSR process used here is a coherence-enhanced nonlinear optical process where the ground state coherence is established by EIT, a resonant atom-light interaction. However, there is still some room for laser detuning optimization, with a minor effect on the squeezing level compared to the zero detuning case. When on resonance, the atomic medium provides the strongest nonlinear response, and our experimental optimization  confirms that squeezing is maximized near zero laser detuning under our primary operating conditions (63.7$^\circ$C, 2 mW/channel). At higher temperature and especially higher laser power, although the nonlinearity is higher, optical pumping also becomes significant, then shifting the laser slightly away from resonance can mitigate resonant absorption while still maintaining an efficient nonlinear interaction for squeezing. As shown in Fig.~\ref{Fig. S19}, the calculated optimal detuning is different for two laser powers of 1 mW/channel and 2 mW/channel, for 70$^\circ$C and 55$^\circ$C. However, the squeezing at optimal red-detuning is only slightly different than at zero-detuning. We note that, since the optimal laser power itself varies with the total atom number through the cell temperature and cell geometry as described above, the optimal detuning also depends on these parameters.

\begin{figure*}
	\centering
	\includegraphics[width=0.5\textwidth]{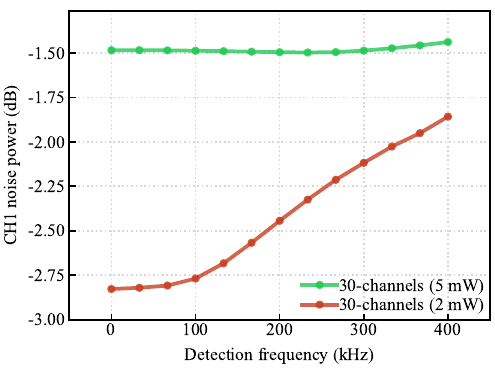}
	\caption{Noise spectrum for different laser power in the array. Numerically calculated squeezed-quadrature noise spectrum for CH1 in a 30-channel array, shown for per-channel optical powers of 2 mW (red) and 5 mW (green). The laser is on resonance with the $|5S_{1/2}, F{=}2\rangle \rightarrow |5P_{1/2}, F'{=}1\rangle$ transition. Noise power is normalized to the photon shot noise level (0 dB). Cell length: 7.5 cm; diameter: 2.5 cm; temperature: 63.7$^\circ$C. It can be seen that there is higher squeezing at lower detection frequency.}
	\label{Fig. S20}
\end{figure*}

\begin{figure*}
	\centering
	\includegraphics[width=0.7\textwidth]{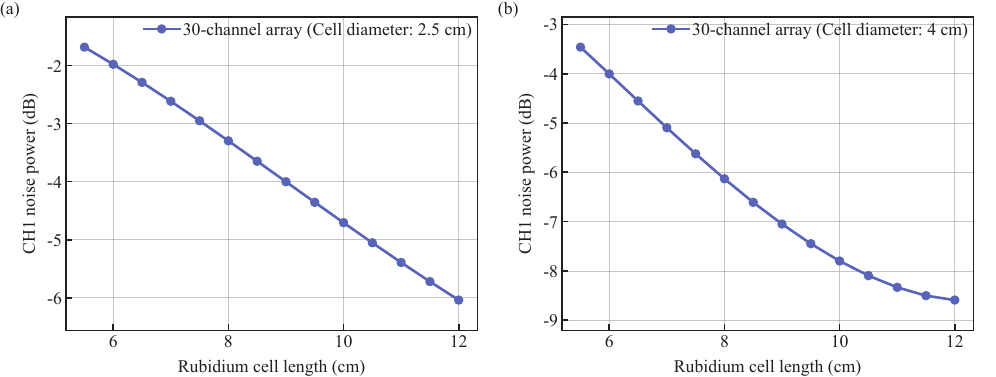}
	\caption{Significantly improved array squeezing results for higher temperature and larger vapor cell. Numerical simulation for the squeezed quadrature noise power of CH1 in a 30-channel array as a function of the length of vapor cell at temperatures of $70$$^\circ$C, where each channel in the array has the same input power of $5$ mW (Rabi frequency of $98.9$ MHz). The laser frequency is red-detuned by $120$ MHz from $|5S_{1/2}F=2\rangle \rightarrow |5P_{1/2}F'=1\rangle$. The detection frequency is fixed at $60$ kHz, and the noise power is normalized to the photon shot noise level ($0$ dB level). (a) Cell diameter: 2.5 cm. (b) Cell diameter: 4 cm.}
	\label{Fig. S15}
\end{figure*}

Another parameter affecting the obtained squeezing is the detection frequency, given a finite bandwidth for light squeezing in general. In our scheme, due to the motion of the atoms it is mainly the array's total laser power averaged across the cell cross-section that determines the squeezing bandwidth. Therefore, theoretically, there is higher squeezing at lower detection frequency, and when we make quantitative comparisons of squeezing under various conditions, the detection frequency is specified. Since the optimal laser power depends on the total number of atoms via the cell geometry and temperature, the squeezing spectrum varies correspondingly with these conditions. Figure.~\ref{Fig. S20} shows calculated squeezing spectra at various optical powers for 63.7$^\circ$C, illustrating the change of squeezing bandwidth with the total laser power. In the present work, experimentally, the detection frequency was chosen near 180 kHz due to dominate technical noises at lower frequencies such as magnetic field noises coupled to the atoms. Importantly, when the cell becomes wider, due to motional averaging of the atoms, the average laser intensity decreases, leading to narrowed squeezing bandwidth. In this case, squeezing detected at lower frequency can be significantly higher than that at higher frequency, which is why the detection frequency is chosen to be 60 kHz in Fig. \ref{Fig. S15} when showcasing improved squeezing.

Building upon the above analysis of how squeezing depends on various parameters, we next showcase the simulation results of significantly improved array squeezing under certain sets of parameters of operating temperature, cell geometry (including cell length and width), laser power, laser detuning, and detection frequency. As shown in Fig. \ref{Fig. S15}, at 70$^\circ$C, by increasing the size of the vapor cell, in particular by using a longer cell and a larger cell cross section, the achievable squeezing of the 30-channel array can exceed 8 dB. We note that, the results in Fig. \ref{Fig. S15} are obtained at a fixed per-channel power for all cell sizes in order to isolate the effect of cell geometry, even though that power may not be optimal operating power for all cell sizes. In a larger cell, the atom number participating in the nonlinear interaction is higher, raising the saturation threshold to higher laser power. Consequently, a condition that is near-optimal for a larger cell may correspond to a significantly sub-saturated regime for a smaller cell, and vice versa. Therefore, jointly optimizing the cell temperature (including the coating material), cell geometry and laser parameters represent an effective and practical route to substantially enhancing array squeezing beyond the levels demonstrated in our current experiment.

\bibliographystyle{apsrev4-2}  
\bibliography{reference_v2}   
	
\end{document}